\documentclass[12pt]{article}
\usepackage{amssymb}
\usepackage{amsmath}
\usepackage{graphicx}
\usepackage{sint,cite}

\newcommand\be{\begin{equation}}
\newcommand\ee{\end{equation}}
\newcommand\bea{\begin{eqnarray}}
\newcommand\eea{\end{eqnarray}}
\newcommand\ba{\begin{array}}
\newcommand\ea{\end{array}}

\def\Im{{\rm Im}\,}
\def\Re{{\rm Re}\,}
\def\Tr{{\rm Tr}\,}
\newcommand{\eq}[1]{Eq.~(\ref{#1})}
\newcommand{\fig}[1]{Fig.~\ref{#1}}
\newcommand{\tab}[1]{Table~\ref{#1}}
\newcommand{\sect}[1]{Section~\ref{#1}}
\newcommand{\ev}[1]{\left\langle #1 \right\rangle}

\def\MeV{\,{\rm MeV}}
\def\fm{\,{\rm fm}}
\def\Nf{N_{\rm f}}

\newcommand{\onecol}[2]{
        \begin{minipage}[t]{#1}{#2\vfill} \end{minipage}
        }

\def\Up{U^\prime}

\def\hP{\hat{P}}
\def\Deta{D[\eta]}
\def\Upp{U^{\prime\prime}}
\newcommand{\pacc}{P_{\rm acc}}
\newcommand{\Pacc}{P_{\rm acc}}
\newcommand{\minm}[1]{\min\left\{1,#1\right\}}
\def\hl{\hat{\lambda}}
\def\varhl{\sigma^2_{\hl}}

\newcommand{\mc}[3]{\multicolumn{#1}{|#2}{#3}}
\def\Sw{S_{\rm w}}
\def\Dw{D_{\rm w}}
\def\mps{m_{\rm PS}}
\def\N1{\bar{N}_1}

\def\ma[#1,#2,#3,#4]  {{\left( \matrix{ #1  & #2 \cr
                                        #3  & #4 \cr } \right)}}

\begin{document}

\thispagestyle{empty}
\title{{\normalsize
\mbox{} \hfill
\onecol{4.0cm}{\vspace{-1.9cm} WUB/12-04}} \\
\vspace{1cm}
Fermions as Global Correction: the QCD Case
}

\author{\normalsize
Jacob Finkenrath $^{a}$, Francesco Knechtli $^{a}$, Bj\"orn Leder $^{a,b}$\\[0.5cm] \normalsize
$^{a}$ Department of Physics, Bergische Universit\"at Wuppertal\\\normalsize
Gaussstr. 20, D-42119 Wuppertal, Germany\\[0.25cm]\normalsize
$^{b}$ Department of Mathematics, Bergische Universit\"at Wuppertal\\\normalsize
Gaussstr. 20, D-42119 Wuppertal, Germany\\[0.5cm]
}
\date{}

\maketitle

\begin{abstract}
It is widely believed that the fermion
determinant cannot be treated in global acceptance-rejection steps of gauge
link configurations that differ in a large fraction of the links. However, for 
exact factorizations of the determinant that separate the ultraviolet from 
the infrared modes of the Dirac operator it is known that the latter show
less variation under changes of the gauge field compared to the former. 
Using a factorization based on recursive domain decomposition allows for a 
hierarchical algorithm that starts with pure gauge updates of the links 
within the domains and ends after a number of filters with a 
global acceptance-rejection step. Ratios of determinants have to be treated 
stochastically and we construct techniques to reduce the noise.
We find that the global acceptance rate
is high on moderate lattice sizes and demonstrate the effectiveness
of the hierarchical filter.
\end{abstract}

%\begin{flushright}
%\end{flushright}

\newpage

\section{Introduction \label{s_intro}}

The state of the art simulation algorithm for lattice QCD is the
Hybrid Monte Carlo (HMC) \cite{Duane:1987de,Gottlieb:1987mq}.
As the continuum limit is approached, when the lattice spacing $a$ goes to zero, 
the simulation cost for a given observable scales typically as $a^{-(5+z)}$.
The dynamical critical exponent $z$ depends on the observable and is
responsible for the critical slowing down of the simulations. Recently in
\cite{Schaefer:2010hu} it was shown that $z(Q^2)=5$ for the 
topological charge $Q$ (the scaling might even be exponential in $1/a$ 
cf. \cite{DelDebbio:2002xa}). This is a common problem for all
present algorithms for gauge theories and the reason has been traced back to
the fact that simulations on periodic lattices get stuck in
topological sectors \cite{Luscher:2010iy}. In fact, on lattices with open
boundary conditions $z(Q^2)=2$ is found in \cite{Luscher:2011kk}. Our original
motivation was to look for an alternative algorithm which allows for 
larger steps in the space of gauge fields.

In recent years new actions to simulate QCD on the lattice have been developed,
in particular, based on smearing of the gauge links in the Dirac operator.
In the case of Wilson fermions the stability of the HMC algorithm is influenced
by the fluctuations of the smallest eigenvalues of the Wilson--Dirac operator
\cite{DelDebbio:2005qa}. The results of \cite{Durr:2010aw} show evidence
that smearing improves the stability. 
The HMC requires the computation of forces (i.e. derivatives of the Dirac operator
with respect to the gauge links) and this can be very complicated or 
even impossible, like when HYP smearing \cite{Hasenfratz:2007rf} is used.
A number of solutions exist, like using stout \cite{Morningstar:2003gk},
nHYP \cite{Hasenfratz:2007rf} or HEX \cite{Capitani:2006ni,Durr:2010aw}
smearing or a differentiable approximation to the SU(3) projection for
the smeared links \cite{Kamleh:2004xk}, but
flexibility in the choice of gauge and fermion actions is highly desirable 
and so the question arises, whether an alternative algorithm without force computations exists.

In this article we study, motivated by a previous work in the Schwinger model
\cite{Knechtli:2003yt}, an algorithm based on global acceptance-rejection steps 
accounting for the fermion determinant in QCD with $\Nf=2$ quark flavors.
The basic idea is to make a gauge proposal which is accepted with a probability
that depends on the ratio of fermion determinants on the ``new'' and ``old''
gauge configurations.
Such an algorithm 
has already been used in QCD simulations with HYP-smeared link staggered
fermions \cite{Hasenfratz:2002jn,Hasenfratz:2002pt,Hasenfratz:2002vv}, with
the fixed point action \cite{Hasenfratz:2005tt} and in \cite{Joo:2001bz}.
The problem with this type of algorithms is their scaling with the lattice
volume $V$.\footnote{Unless otherwise specified, in this article we use lattice 
units (i.e. we set $a=1$) and $V$ is the number of lattice points.}
The cost of an exact determinant computation grows with $V^3$ and the 
acceptance to change a finite fraction of links decreases like $\exp{(-V)}$.

In order to avoid the computation of exact determinants we use a
stochastic estimation. This estimation can naively introduce a noise which
grows like $\exp{(V)}$.
In order to tackle these problems we construct a hierarchical filter of acceptance-rejection steps
which successively filters the large fluctuations of the gauge proposal \cite{Finkenrath:2011py}. 
Hierarchical acceptance-rejection steps based on approximations of the
determinant with increasing accuracy were introduced and tested in
\cite{Hasenbusch:1998yb}. Here the filter relies on an exact factorization of
the fermion determinant based on domain decomposition \cite{Luscher:2005rx},
which separates the short distance
from the long distance scales of the lattice.
A hierarchy of block acceptance-rejection steps was proposed in
\cite{Luscher:2003vf} but has never been tested.

The article is organized as follows. In \sect{s_hierarchy} we introduce
the hierarchical filter of acceptance-rejection steps. Its construction
based on domain decomposition is detailed in \sect{s_dd}. The techniques
we use for the stochastic estimation of determinant ratios are presented in
\sect{s_techn}. In particular we introduce an interpolation of the gauge
fields which also allows to compute the exact (i.e. without stochastic noise)
acceptance. Results for the latter and the effectiveness of the filter
are shown in \sect{s_exactacc}. \sect{s_res} presents simulation results
of $16^4$ and $32\times16^3$ lattices using a filter with three 
acceptance-rejection steps. A comparison with the HMC is made
for observables like the plaquette or the topological charge.
In the conclusions \sect{s_concl} we also discuss the scaling with the
volume. Appendix \ref{s_appa} contains the proof of detailed balance,
Appendix \ref{s_appb} describes the technique of relative gauge fixing
used for the stochastic estimation and Appendix \ref{s_appc} explains
how the acceptance is enhanced by the use of additional parameters.

\section{Hierarchy of acceptance steps \label{s_hierarchy}}

Let $P(s)$ be the desired distribution of the states $s$ of a system. Suppose a
process that proposes a new state $s'$ with transition probability 
$T_0(s\to s')$ and 
fulfills detailed balance with respect to $P_0(s)$. A process with fixed point
distribution $P(s)$ is then obtained by the combination of such a proposal with
a subsequent Metropolis acceptance-rejection step \cite{Metropolis:1953am}
\begin{equation}
\begin{split}
 0) &\quad \text{Propose $s'$ according to $T_0(s\to s')$}\\
 1) &\quad \Pacc(s\to s') = \minm{\frac{P_0(s)P(s')}{P(s)P_0(s')}} \,.
\end{split}
\end{equation}
This hierarchy of a proposal step and an acceptance-rejection step
can easily be generalized to an arbitrary number of acceptance-rejection
steps. The result of the first acceptance-rejection step $1)$ is then
interpreted as the proposal for a second acceptance-rejection step $2)$ and so
on. If the target distribution $P(s)$ factorizes into $n+1$ parts
\begin{equation}\label{eq:factor}
 P(s)=P_0(s)\,P_1(s)\,P_2(s)\dots P_{n}(s)\,,
\end{equation}
the resulting hierarchical acceptance-rejection steps take the form
\begin{equation}\label{eq:hierarchy}
\begin{split}
 0) &\quad \text{Propose $s'$ according to $T_0(s\to s')$} \\
 1) &\quad \Pacc^{(1)}(s\to s') = \minm{\frac{P_1(s')}{P_1(s)}} \\
 2) &\quad \Pacc^{(2)}(s\to s') = \minm{\frac{P_2(s')}{P_2(s)}} \\
 ...\\
 n) &\quad \Pacc^{(n)}(s\to s') = \minm{\frac{P_n(s')}{P_n(s)}} \,.
\end{split}
\end{equation}
In the context of lattice QCD it is plausible to assume
$P_i(s)\propto\exp(-S_i(s))$ with real actions $S_i$ and thus
\be
\frac{P_i(s')}{P_i(s)} = {\rm e}^{-\Delta_i(s,s')}\,,
\ee
where $\Delta_i(s,s')=S_i(s')-S_i(s)$. 
The average acceptance rate in step $i)$ is defined by
\be
 \ev{\Pacc^{(i)}}_{s,s'} = \sum_s P(s)\sum_{s'} P_0(s')\,P_1(s')\dots P_{i-1}(s')\,
 \minm{{\rm e}^{-\Delta_i(s,s')}}\,.
\ee
It can be computed assuming a Gaussian distribution for $\Delta_i(s,s')$ with
variance $\Sigma_i^2$ and the result is \cite{Knechtli:2003yt} (see also 
\cite{Irving:1996bm})
\begin{equation}\label{eq:acceptance}
 \ev{\Pacc^{(i)}}_{s,s'} = {\rm erfc}\left(\sqrt{\Sigma_i^2/8}\right)\,.
\end{equation}
The acceptance rates might be enhanced by parameterizing and tuning the
factorization \eqref{eq:factor}, see Appendix \ref{s_appc}.

Our goal is to simulate QCD with $\Nf=2$ mass-degenerate fermions.
After integration over the Grassmann fermion
fields the states $s$ are defined by the gauge field $U$ and
the target probability distribution is
\be
P(U) = \frac{\left|\det(D(U))\right|^2{\rm e}^{-S_g(U)}}{Z} \,, \label{2flavordistr}
\ee
where $S_g$ is the gauge action, $D$ is the lattice Dirac operator and $Z$ is
the partition function
\be
Z = \int D[U] |\det D(U)|^2 {\rm e}^{-S_g(U)} \,.
\ee
The integration measure is
$D[U] = \prod_{x,\mu} dU(x,\mu)$, where
$dU(x,\mu)$ is the SU(3) Haar measure for the link $U(x,\mu)$.

A simple two-step algorithm would consist of some update
of the gauge link configuration $U\to \Up$, which fulfills
detailed balance with respect to $P_0(U)\propto \exp(-S_g(U))$, followed by an
acceptance-rejection step with the fermion determinant ratio
\begin{equation}
 \Pacc^{(1)}(U\to \Up) = \minm{\det 
 \frac{D(\Up)^\dagger D(\Up)}{D(U)^\dagger D(U)}} \,. \label{paccdetratio}
\end{equation}
The proof of detailed balance can be found in \sect{2stepex}.

If the proposal changes only one link and the Dirac operator $D$ is
ultra-local it is easy to show that the acceptance-rejection step requires
only few inversions\footnote{For example, in the case of the Wilson--Dirac
operator 12 inversions are needed.}.
An ergodic algorithm is then obtained by sweeps
through the lattice. Thus the cost of such an algorithm would scale 
with the lattice volume $V$ at least
like $V^2$ \cite{Weingarten:1980hx} and it requires O($V$) inversions per
sweep.

If, on the other hand, a finite fraction $\propto V$ of the links is updated
for the proposal, the acceptance rate decreases exponentially with the volume.
In order to see this we write the distribution $P_1$ as
$P_1(U)\propto\exp(\ln(\det\, D^\dagger D))$. 
The action difference $\Delta_1(U,\Up)=
\ln(\det\, D(\Up)^\dagger D(\Up)) - \ln(\det\, D^\dagger(U) D(U))$ can
be written as $\Delta_1=-\sum_i\ln(\lambda_i)$
in terms of the eigenvalues $\lambda$ of the operator $M^\dagger M$
with
\be
M = D(\Up)^{-1}D(U) \,. \label{Mdef}
\ee
If we assume a Gaussian distribution\footnote{
We verified numerically that this assumption is valid to a good
approximation.} (after averaging over the gauge ensemble
$U$ and the proposals $\Up$) for the logarithms of the
eigenvalues $\hl_i=\ln(\lambda_i)$ with mean zero and variance 
$\varhl$, we can approximate
\be
\Sigma_1^2 \approx \N1\,\varhl/2 \,, \label{varexapprox}
\ee
where $\N1$ is the number of eigenvalues $\lambda\neq1$. Typically
$\N1\propto V$ and this implies that $\Sigma_1^2$ is
proportional to the volume $V$.
The complementary error function in the formula for the acceptance
\eqref{eq:acceptance} has the asymptotic expansion 
${\rm erfc}(x)\sim\exp(-x^2)/(x\sqrt{\pi})(1-1/(2x^2)+\cdots)$
for $|x|\gg1$
which shows the exponential decrease with the volume.

From the preceding discussion it is obvious that such two-step algorithms will
not be efficient for large lattices. Indeed numerical experiments show that
for lattices larger than $\sim(0.2\;\mathrm{fm})^4$ (where all links are
updated) the acceptance rate quickly becomes less than a percent. However, in
the context of low mode reweighting the fluctuations of the determinant of
$D_{\text{low}}^\dagger D_{\text{low}}$, where $D_{\text{low}}$ is a
restriction of $D$ to its low modes, are found to depend only mildly on the
volume \cite{Luscher:2008tw}. The explanation for this observation might be
the fact that the width of the distribution of the small eigenvalues of 
$\sqrt{D^\dagger D}$ decrease like $1/V$ \cite{Luscher:2008tw} (the fluctuations of the
eigenvalue gap go instead like $1/\sqrt{V}$ \cite{DelDebbio:2005qa}). 
Thus, given a factorization of the determinant that separates low (infrared IR)
and high (ultraviolet UV) modes
\begin{equation}
 \det(D) = \det(D_{\text{UV}}) \cdots \det(D_{\text{IR}})\,,
\end{equation}
a hierarchy of acceptance steps can be constructed, where the large
fluctuations of the UV modes go through a set of filters (acceptance-rejection
steps) which are more and more dominated by the IR modes:
\begin{center}
\begin{tabular}{cccccc}
$0)$ & $P_0$ & UV & short distance & local & cheap\\
& $\vdots$ & $\vdots$ & $\vdots$ & $\vdots$ & $\vdots$ \\
$n$) & $P_n$ & IR & long distance & global & expensive
\end{tabular}
\end{center}
This hierarchy of modes may induce also a hierarchy of costs since it is the
low modes that cause the most cost in lattice QCD. Furthermore the
factorization should be exact and the terms simple to compute. Factorizations
that realize these conditions are already used to speed-up the HMC algorithm,
i.e., in the context of mass-preconditioning \cite{Hasenbusch:2001ne} and
domain decomposition \cite{Luscher:2005rx}. Only the latter also allows for a
decoupling of local updates and will be discussed in the following.

\section{Domain decomposition \label{s_dd}}

Domain decomposition was introduced in lattice QCD in \cite{Luscher:2003vf}
and in \cite{Luscher:2005rx} the
resulting factorization of the fermion determinant was used to separate short
distance and long distance physics in the HMC algorithm.
For definiteness we consider here the Wilson--Dirac operator $D(U)$ 
\cite{Wilson:1974sk}, which may include the clover term needed for
O($a$) improvement \cite{Sheikholeslami:1985ij,Luscher:1996sc}.
But our algorithm
is applicable to a more general class of Dirac operators, see below.
%
%%%%%%%%%%%%%%%%%%%%%%%%%%%%%%%%%%%%%%%%%%%%%%%%%%%%%%%%%%%%%%%%%%%%%%%%%%%%%%%%
\begin{figure}[tb]
 \centering
 \includegraphics[width=0.5\textwidth]{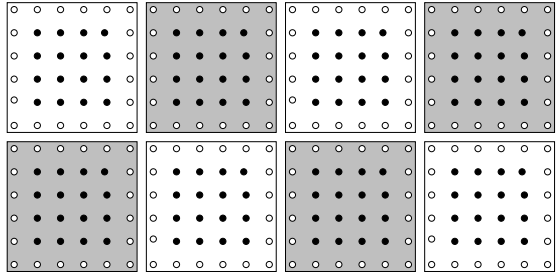}
 \caption{Block decomposition of a 2-dimensional lattice. The blocks are
   coloured like a checker board. Picture taken from \cite{Luscher:2005rx}.}
 \label{fig:bgr}
%\vspace{-1em}
\end{figure}
%%%%%%%%%%%%%%%%%%%%%%%%%%%%%%%%%%%%%%%%%%%%%%%%%%%%%%%%%%%%%%%%%%%%%%%%%%%%%%%%
%

Suppose a decomposition $\mathcal{C}$ of the lattice in non-overlapping blocks $
b\in \mathcal{C}$
(cf.~\fig{fig:bgr} for a 2-dimensional visualization).
The lattice sites are labeled such that the
sites belonging to the first black block come first,
then the second black block and after the last black block the first white block
and so on. The Dirac operator can then be written as
\begin{equation}\label{eq:domdec}
 D = \left(\begin{array}{cc} D_{\rm bb} & D_{\rm bw} \\
D_{\rm wb} & D_{\rm ww} \end{array}\right)\,,
\end{equation}
where $D_{\rm bb}$ ($D_{\rm ww}$) is a block-diagonal matrix with the black
(white) block Dirac operators $D_b$ on the diagonal. The block Dirac operators
$D_b$ fulfill Dirichlet boundary conditions and therefore are dominated by
short distance physics (if the blocks are small enough).
The matrices $D_{\rm bw}$ and $D_{\rm wb}$
contain the block interaction terms. The form \eqref{eq:domdec} induces a
factorization of the determinant
\begin{equation}\label{eq:ddfactor}
 \det(D) = \prod_{b \in \mathcal{C}}\det(D_b)\det(\hat{D})\,, \quad \hat{D} = 1
- D_{\rm bb}^{-1}D_{\rm bw}D_{\rm ww}^{-1}D_{\rm wb}\,,
\end{equation}
where $\hat{D}$
is the Schur complement of the decomposition \eqref{eq:domdec} and contains
block interactions, i.e.~the long distance physics. A natural separation scale
is given by the inverse block size $1/L_b$. 
In the context of the domain decompositioned HMC the average force associated
with the Schur complement is an order of magnitude smaller than the
force associated with the block Dirac operators \cite{Luscher:2005rx}.
This indicates that the fluctuations of the determinant of the Schur
complement are smaller than that of the block determinants. 
Furthermore the factorization
\eqref{eq:ddfactor} can be iterated using a recursive domain decomposition
\begin{equation}\label{eq:ddrec}
 \det(D_b) = \prod_{b' \in \mathcal{C}_b}\det(D_{b'})\det(\hat{D}_b)\,.
\end{equation}
We note that the Schur complement $\hat{D}_b$ fulfills Dirichlet boundary
conditions.
We have implemented the recursive domain decomposition in the
freely available software package DD-HMC by M. L\"uscher \cite{soft:DDHMC}.
In the case of one level of recursion the hierarchy of acceptance-rejection
steps is given by
\begin{equation}\label{eq:4step}
\begin{split}
 1) &\quad \Pacc^{(1)} = \minm{\det \frac{D_{b'}(\Up)^\dagger
     D_{b'}(\Up)}{D_{b'}(U)^\dagger D_{b'}(U)}}\,,\quad \forall b, \forall b'
 \in \mathcal{C}_b \\
 2) &\quad \Pacc^{(2)} = \minm{\det \frac{\hat{D}_b(\Up)^\dagger
     \hat{D}_b(\Up)}{\hat{D}_b(U)^\dagger \hat{D}_b(U)}}\,,\quad \forall b \in
 \mathcal{C} \\
 3) &\quad \Pacc^{(3)} = \minm{\det \frac{\hat{D}(\Up)^\dagger
     \hat{D}(\Up)}{\hat{D}(U)^\dagger \hat{D}(U)}} \,.
\end{split}
\end{equation}
At the beginning the set of links to be updated, the so called
\emph{active} links, is chosen such that
the acceptance-rejection steps for the smallest blocks, $b'$, at stage $1)$
in \eq{eq:4step} decouple and can therefore be processed in parallel.
In the case of Wilson fermions with or without clover term the active links are
the links that have at most one endpoint
on the boundary of a block (white points in \fig{fig:bgr}). In this case the
block acceptance steps also decouple if the links in the
Wilson--Dirac operator (but not in the clover term) are replaced by one level
of HYP smearing \cite{Hasenfratz:2001hp}. After the last and global
acceptance-rejection step the gauge field is translated by a random vector,
see Appendix C of \cite{Luscher:2005rx}.

If the smallest blocks, $b'$, at stage $1)$ in \eq{eq:4step}
consist of no more than $\sim 6^4$ lattice points, the determinant ratios
can be efficiently computed exactly by LU-decomposition \cite{soft:sparseLU}. 
If the smallest blocks
are larger, we compute their determinants by a factorization like in 
\eq{eq:ddrec}. The Schur complements at the stages 
$2)$ and $3)$ in \eq{eq:4step} are usually too large for their determinant
ratios to be computed exactly and have 
to be treated stochastically\footnote{
The same applies to Schur complements arising from a factorization of the
smallest blocks, if that is needed.}.
The stochastic estimation of determinant ratios is the topic of the next
section.
Following this discussion
we give to our algorithm the name of {\em Partially Stochastic Multi-Step}
(PSMS) algorithm.

\section{Stochastic techniques for determinant ratios \label{s_techn}}

Since the numerical cost for the computation of exact determinants
grows with the cube of the size of the matrix,
determinants of Dirac operators for lattices larger than $6^4$ have to
be estimated stochastically. 
In particular for our problem we have
to estimate ratios of determinants of Schur complements, which arise from
a domain decomposition and appear in the acceptance-rejection steps of
\eq{eq:4step}.
In Appendix \ref{s_appa} we show that such stochastic acceptance-rejection
steps fulfill detailed balance.
In this section we describe in detail the techniques we use
to reduce the associated stochastic noise.

In \sect{subs_stochdet} we discuss the stochastic noise introduced
when the determinant ratio in \eq{paccdetratio} (for generic Dirac operators
$D$) is evaluated stochastically. The stochastic noise depends
on the spectrum of generalized eigenvalues of the
operators forming the ratio \cite{Knechtli:2003yt}.
In order to reduce it we apply
techniques described in \sect{subs_relgf} and \sect{subs_gfp}.
In \sect{subs_relgf} we discuss a relative gauge fixing 
of the gauge field $U$ and $\Up$. This gauge fixing is
applied for the construction of a gauge field interpolation, 
a new method which we present in \sect{subs_gfp}.
The gauge fields are linearly interpolated and this
induces a factorization in terms of ratios of operators which can
be made arbitrarily close as the number of interpolation steps increases. 
In particular, there exists the limit in
which the exact ratio is obtained.
In \sect{subs_schur} the properties of
the Schur complement are reviewed. In this particular case the noise
vector can be restricted to a subspace of the boundary points of the blocks.
In \sect{subs_technnum} we support the introduction of these techniques
by numerical results.

\subsection{Stochastic estimation of determinant ratios \label{subs_stochdet}}

We replace
the determinants of ratios of Dirac operators in \eq{paccdetratio}
by stochastic estimators
\begin{equation}
\minm{\det(M^\dagger M)^{-1}} \longrightarrow \minm{{\rm
    e}^{-|M\eta|^2+|\eta|^2}}\,,\label{stacc}
\end{equation}
where the ratio operator $M$ is defined in \eq{Mdef}.
In \eq{stacc} $\eta$ is a complex Gaussian noise vector that is updated before
each acceptance-rejection step and $|\eta|^2$ is its norm squared, see
\sect{s_stoch}. The average over $\eta$ of a function $f(\eta)$ is defined by
\be
\ev{f(\eta)}_\eta = \int\Deta\,{\rm e}^{-|\eta|^2}f(\eta) \,. \label{aveta}
\ee
The measure $\Deta$ is normalized such that $\int\Deta\,\exp(-|\eta|^2)=1$.
The algorithm satisfies detailed balance (the proof is given in \sect{s_stoch})
and yields an acceptance rate that is bounded from above by the exact 
acceptance in \eq{paccdetratio} \cite{Knechtli:2003yt}.
There are other possible choices for the distribution of $\eta$
than a Gaussian distribution. But because of the central limit
theorem these other choices are equivalent to the Gaussian distribution in the
large volume limit.

The stochastic noise introduced in the acceptance-rejection step by \eq{stacc}
has the effect of replacing in \eq{eq:acceptance}
\be
\Sigma^2 \longrightarrow 
\sigma^2 = \Sigma^2+\left(\sigma^{\rm stoch}\right)^2
\,, \label{varexactplusstoch}
\ee
where
\be
\left(\sigma^{\rm stoch}\right)^2 =
\ev{\Delta^2}_{U,\Up,\eta} - (\ev{\Delta}_{U,\Up,\eta})^2  \label{varstoch}
\ee
with
$\Delta = |M\eta|^2 - |\eta|^2$. The average $\ev{\cdot}_{U,\Up,\eta}$ is taken over the gauge ensemble
$U$, the proposals $\Up$ and the noise vectors $\eta$. For given $U$ and $\Up$
\eq{varstoch} can be computed by performing the integrations over
$\eta$ in the basis of orthonormal eigenvectors of $M^\dagger M$ with
eigenvalues\footnote{
The eigenvalues $\lambda$ of $M^\dagger M$ are equivalent to the generalized
eigenvalues of the problem $D(U)D(U)^\dagger\chi=\lambda D(\Up)D(\Up)^\dagger\chi$.} 
$\lambda_k$, cf. \cite{Knechtli:2003yt}. The result is
\be
\left(\sigma^{\rm stoch}\right)^2 = \ev{\sum_k (\lambda_k-1)^2}_{U,\Up} \,. \label{varstoch2}
\ee
The eigenvalues $\lambda=1$ do not contribute to the variance. If we denote
by $h_1$ the full width at half maximum (FWHM) of the distribution of the
eigenvalues
$\lambda_k$ and by $\N1$ the number of eigenvalues which are not one,
we can approximate
\be
\left(\sigma^{\rm stoch}\right)^2 \approx \N1\,h_1^2 \,. 
\label{varstochapprox}
\ee
It becomes clear that the smaller $\N1$ and $h_1$ are, the larger the
stochastic acceptance will be. Furthermore in \cite{Hasenfratz:2002ym} it
was noted that the spectrum of $M^\dagger M$ has to fulfill the condition
$\lambda>0.5$, because otherwise the variance of the quantity under the
minimum function in \eq{stacc} is not defined.

\subsection{Relative gauge fixing \label{subs_relgf}}

In \cite{Knechtli:2003yt} (see also \cite{Hasenfratz:2005tt}) it was noticed
that relative gauge fixing of the configuration $U$ and $\Up$ reduces the
stochastic noise in \eq{varstoch}.
Under a gauge transformation $g(x)\in\,{\rm SU(3)}$, the gauge links transform
as $U(x,\mu)\to U^g(x,\mu)=g(x)U(x,\mu)g(x+\hat{\mu})^{-1}$
and the Dirac operator as
\be
D(U^g)_{xy} = g(x) D(U)_{xy} g(y)^{-1} \qquad \mbox{(no sum over $x$ and $y$)}\,,
\ee
where we suppress the spin indices.
Further we define a scalar product of two gauge fields as
\be
(U,\Up) = \frac{1}{12V}\sum_{x,\mu}\Re\Tr
         \left\{1-U(x,\mu)^\dagger \Up(x,\mu)\right\} \,. \label{sprodgf}
\ee
Relative gauge fixing is defined through the minimization
\bea
&& \min_{g_1,g_2}\, (U^{g_1},{\Up}^{g_2}) = 
\min_{g_1,g_2} \frac{1}{12V}\sum_{x,\mu}\Re\Tr\Big\{
\nonumber \\
&& 1-g_1(x+\hat{\mu})U(x,\mu)^\dagger g_1(x)^{-1}
g_2(x)\Up(x,\mu)g_2(x+\hat{\mu})^{-1}\Big\} \,. \label{relgf}
\eea
We determine $g_1$ and $g_2$ before the acceptance-rejection step \eq{stacc}, where we use
\be
M = D({\Up}^{g_2})^{-1}D(U^{g_1}) \,. \label{matrixMgf}
\ee
Relative gauge fixing does not change the exact acceptance rates in \eq{eq:4step} but
in general improves the stochastic acceptance rate in \eq{stacc}. In order to show
detailed balance in the latter case, consider the reverse transition
$\Up\to U$, for which the minimization is 
$\min_{\tilde{g}_1,\tilde{g}_2}\, ({\Up}^{\tilde{g}_1},U^{\tilde{g}_2})$.
As one can immediately see by taking the complex conjugate of \eq{relgf}
the result is given by $\tilde{g}_1=g_2$ and $\tilde{g}_2=g_1$. This implies for the
reverse transition
\be
M\to
\tilde{M} = D(U^{\tilde{g}_2})^{-1}D({\Up}^{\tilde{g}_1}) = M^{-1} \,,
\ee
which is precisely the property needed to prove detailed balance
\cite{Knechtli:2003yt}.

In the above procedure, the choice of $g_1$ and $g_2$ is not unique. In fact
one can transform $g_1\to g_1h$ and $g_2\to g_2h$ by
some other gauge transformation $h(x)$ and the minimization condition
\eq{relgf} is unchanged. Instead we choose\footnote{
We thank Ulli Wolff for suggesting this choice.}
\be
  g_1 = g_2^{-1} = g \,. \label{choiceg}
\ee
The numerical procedure for the minimization \eq{relgf} using \eq{choiceg}
is described in Appendix \ref{s_appb}.

In the proposal $U\to\Up$ we only change active links in the blocks and we
restrict the gauge transformations $g$ in \eq{choiceg} to the black points
 in \fig{fig:bgr}. One reason for this is that
the critical slowing down of such a local (i.e. restricted to the blocks)
minimization is reduced compared to a global minimization over the entire
lattice. 

\subsection{Gauge field interpolation \label{subs_gfp}}

In order to ensure $\lambda>0.5$ and bring the spectrum 
of $M^\dagger M$ closer to one, one could employ the method of determinant
breakup introduced in \cite{Hasenbusch:1998yb,Hasenfratz:2002ym}.
It uses the factorization $\det(M^\dagger M)= [\det((M^\dagger M)^{1/N})]^N$ and
in the stochastic acceptance-rejection step \eq{stacc} each factor is then 
replaced by a stochastic estimator with an independent noise vector.
The effect on the spectrum of $M^\dagger M$ is to replace
$\lambda\to\lambda^{1/N}$. The gauge field interpolation which we propose
in this article has a similar
effect but avoids the computation of $1/N$th roots of $M^\dagger M$.

We introduce a sequence of intermediate fields
$U_i,\;i=0,\ldots,N$ which starts from the gauge field $U_0=U^{g}$ and 
ends with the gauge field $U_N={\Up}^{g^{-1}}$.
$g$ is the gauge transformation in \eq{choiceg}.
The determinant of $M^\dagger M$ can be factorized like
\begin{equation}\label{eq:param}
\det(M^\dagger M) = \prod_{i=0}^{N-1} \det(M_i^\dagger M_i)\,,
\end{equation}
where
\be
M_i = D(U_{i+1})^{-1} D(U_i) \,.\label{Mparam}
\ee
The stochastic acceptance-rejection step in \eq{stacc} is done
by drawing one independent Gaussian noise vector $\xi_i$ for
each factor
\begin{equation}
\minm{{\rm e}^{\sum_{i=0}^{N-1}-|M_i\xi_i|^2+|\xi_i|^2}}\,.\label{staccN}
\end{equation}
The cost is then one inversion for each factor.
In order for the algorithm to fulfill detailed balance the intermediate gauge
configurations have to be the same when doing the reverse change $\Up\to
U$. The proof of detailed is given in \sect{s_stochpar}.

The simplest way to construct such an interpolation is
\be
U_i(x,\mu) = \frac{N-i}{N} U^{g}(x,\mu) + \frac{i}{N} {\Up}^{g^{-1}}(x,\mu) \,, \quad i=0,1,\cdots,N-1 \,,
\label{gfparam}
\ee
which interpolates linearly between $U_0=U^{g}$ and $U_N={\Up}^{g^{-1}}$.
The interpolation has no physical meaning, only
numerical efficiency counts. 
The intermediate fields are not SU(3) matrices, in the Dirac operator we use
$U_i^{\dagger}$ (and not $U_i^{-1}$) in order to preserve the $\gamma_5$ Hermiticity
of the Wilson--Dirac operator.
Since \mbox{$||U_i- U_{i+1}||\propto 1/N$},
$\forall\, i<N$, we expect the eigenvalues $\lambda_k^{(i)}$ of $M_i^\dagger
M_i$ to be \mbox{$\lambda_k^{(i)}=1+{\rm O}(h_1/N)$} and so the FWHM
of their distribution\footnote{
In the case of the full Dirac operator, we find numerically that
the smallest (largest) eigenvalue change with $N$ as
$\lambda^{(i)}_{\rm min}\sim \exp\{-b/N\}$ ($\lambda^{(i)}_{\rm max}\sim
\exp\{b'/N\}$) for positive constants $b$ ($b'$). This is the same behavior
one obtains using the determinant breakup in $1/N$th roots.}
 can be approximated by $h_N\approx h_1/N$ 
in terms of the FWHM $h_1$ of the eigenvalue distribution of $M^\dagger M$.
The stochastic noise in the acceptance-rejection step is reduced to 
\be
\left(\sigma_N^{\rm stoch}\right)^2 \approx N\,\N1\,h_N^2 \approx
\N1\,\frac{h_1^2}{N} \label{varstochapproxN}
\ee
as compared to \eq{varstochapprox}.
An important feature of this method is the limit \mbox{$N\to\infty$},
for which $\sigma_N^{\rm stoch}\to0$ and we recover the {\em exact}
acceptance, cf. \eq{varexactplusstoch}.

\subsection{Schur Complement \label{subs_schur}}

The Schur complement in \eq{eq:ddfactor} is
$\hat{D}=1-Q$ with
$Q=D_{\rm bb}^{-1}D_{\rm bw}D_{\rm ww}^{-1}D_{\rm wb}$.
Let us denote by $P$ the orthonormal
projector to the space of the white points in the black blocks in
\fig{fig:bgr}. For the points which have only one nearest neighbor on a
different block, $P$ projects to only two of the four spin components.
The explicit definition of $P$ can be found in Appendix B of
\cite{Luscher:2005rx}. It does not depend on the gauge field and it
satisfies the properties $D_{\rm wb}P=D_{\rm wb}$ and $P^2=P$ which imply
\be
\det(1-Q) = \det(1-PQ) \,. \label{detschur}
\ee
This means
that one can use $1-PQ$ instead of $\hat{D}$ in \eq{stacc} and therefore
the noise $\eta$ is defined only on the space invariant under $P$.
We also need to apply the inverse of the operator $1-PQ$ which is 
\cite{Luscher:2005rx} 
\be
(1-PQ)^{-1} = 1-PD^{-1}D_{\rm wb} \,.\label{inverseschur}
\ee
Here $D_{\rm wb}$ is meant to act on the total space of points (by padding with zeros).
For a global lattice of sizes $L_\mu$ in
directions $\mu=0,1,2,3$ and a domain decomposition into blocks of sizes
$l_\mu$, the dimension of the space invariant under $P$ is
\be
{\rm dim}(P) =
6\prod_{\mu=0}^3\frac{L_\mu}{l_\mu}\left(
\sum_{\nu=0}^3\frac{l_0l_1l_2l_3}{l_\nu}-4\sum_{\nu=0}^3(l_\nu-1) \right) \,.
\label{dimproj}
\ee
For the number $\N1$ in \eq{varstochapprox} we have
$\N1\le{\rm dim}(P)$.
On a lattice with the same number of points $L$ in all directions, if we
choose $l_\mu=L/2$ (16 blocks) then ${\rm dim}(P)\approx48L^3$, to be compared
to $V=12L^4$ if we were to consider the full Dirac operator.

The reduction of $\N1$ alone
turns out not to be sufficient to make stochastic acceptance-rejection
steps like in \eq{stacc}, with the Schur complement ratio, efficient. 
Moreover the relative gauge fixing described in \sect{subs_relgf} does not
directly help in reducing the stochastic noise in this case.
The reason is that the
restriction of the gauge transformations to the black points in
\fig{fig:bgr} leaves the Schur complement invariant. This is why the
gauge field interpolation is necessary to further reduce the noise. As we show
in the next section relative gauge fixing has an impact on the interpolation.

\subsection{Numerical results \label{subs_technnum}}
%
%%%%%%%%%%%%%%%%%%%%%%%%%%%%%%%%%%%%%%%%%%%%%%%%%%%%%%%%%%%%%%%%%%%%%%%%%%%%%%%
\begin{figure}[t]
 \begin{center}
     \includegraphics[width=0.60\textwidth]{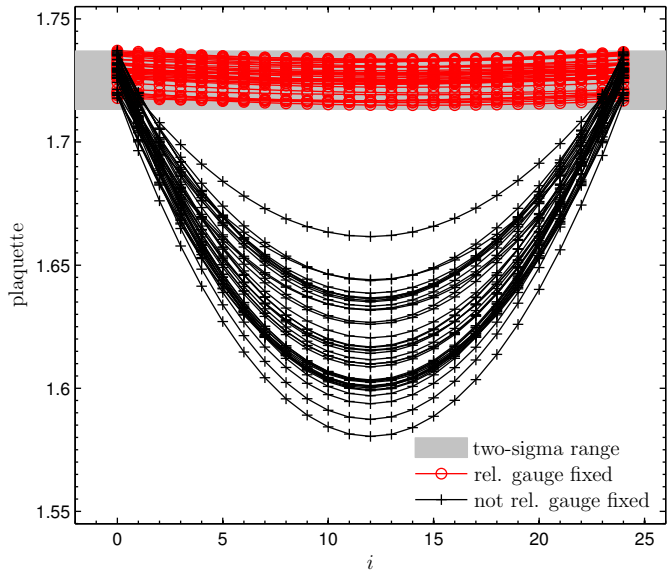}
 \end{center}
% \vspace{-0.5cm}
 \caption{\small The plaquette value for the interpolated 
    fields $U_i$ defined in \eq{gfparam} is shown as a function of $i$.
    The start and end fields $U_0$ and $U_N$ ($N=24$, 40 pairs) are 
    $8^4$ gauge configurations taken from simulations of plain Wilson fermions at
    $\beta=5.6$, $\kappa=0.15825$, where active links in $4^4$ blocks
    are changed.
    We compare plaquette values with (red circles) and without (black pluses)
    relative gauge fixing.
}
 \label{f_plaqparam}
\end{figure}
%%%%%%%%%%%%%%%%%%%%%%%%%%%%%%%%%%%%%%%%%%%%%%%%%%%%%%%%%%%%%%%%%%%%%%%%%%%%%%%
%

\begin{figure}[t]
 \begin{center}
     \includegraphics[width=0.60\textwidth]{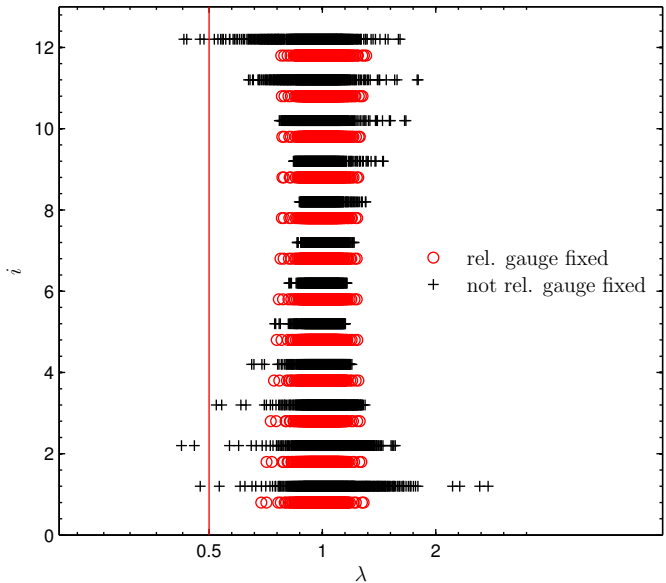}
 \end{center}
% \vspace{-0.5cm}
 \caption{\small The spectrum of the Schur complement ratio
 $M_i^\dagger M_i$ defined in \eq{Mparam} is shown as a function of $i$.
 The start and end fields $U_0$ and $U_N$ ($N=12$) are quenched 
 $4^4$ lattices where all links are changed (and relative gauge fixed).
 The plain Wilson--Dirac operator with mass $am_0=0.56$ is used and the Schur
 complement is defined for a domain decomposition in $2^4$ blocks.
}
 \label{f_schurparam}
\end{figure}

The interpolated fields $U_i$ in \eq{gfparam} change if we apply first a
relative gauge fixing of $U$ and $\Up$, which minimizes their distance in the
sense of \eq{sprodgf}.
In \fig{f_plaqparam} we show the behavior of the plaquette of the interpolated
fields $U_i$. In the computation of the plaquette, if $U$ denotes a link then
the link in reversed direction is defined by $U^{\dagger}$ (and not by $U^{-1}$).
Without relative gauge fixing the intermediate configurations
look like if they were thermalized configurations of a smaller $\beta$. 
The links become rougher. This is understandable if one imagines that the 
gauge configurations $U$ and $\Up$ lay somewhere randomly 
in the configuration space. So the path will not go over configurations which
are similar
to the ``thermalized'' ones.
With relative gauge fixing, the path of the interpolated links yields plaquette
values which are approximately constant, cf. \fig{f_plaqparam} which also shows
the two-sigma band of a thermalized ensemble.
In \fig{f_schurparam} we show the spectra of the Schur complement ratios 
$M_i^\dagger M_i$ in \eq{Mparam}.
Since relative gauge fixing is applied to all links the spectrum is narrower
and the requirement $\lambda>0.5$ can be fulfilled
for a relatively low value of interpolation steps $N$. As expected from the
behavior of the plaquette the width of the spectrum does not change significantly
along the interpolation.
%
%%%%%%%%%%%%%%%%%%%%%%%%%%%%%%%%%%%%%%%%%%%%%%%%%%%%%%%%%%%%%%%%%%%%%%%%%%%%%%%
%%%%%%%%%%%%%%%%%%%%%%%%%%%%%%%%%%%%%%%%%%%%%%%%%%%%%%%%%%%%%%%%%%%%%%%%%%%%%%%
%

There are many possible ways to define alternative interpolations
replacing \eq{gfparam}. For example we could normalize the links by
substituting $U_i(x,\mu)\to U_i(x,\mu)\det(U_i(x,\mu))^{-1/3}$. It turns
out that in this case the intermediate configurations look like if they were
thermalized configurations at larger $\beta$. As a consequence the spectrum 
can develop negative eigenvalues for small quark masses.
We note that a mass-shift towards larger
masses can be generated by multiplying the links with a common 
factor $\exp(\alpha)$, $\alpha<0$, in the Dirac operator. 
Effectively such a factor (albeit with a different value for each link)
can be easily 
incorporated into \eq{gfparam} by multiplying the links $U_i(x,\mu)$ with
an appropriate power of their determinant $\det(U_i(x,\mu))$.
If we do not normalize the links, a ``mass shift'' towards larger
masses is automatically realized because $\det(U_i(x,\mu))<1$.
But there is some room for improving the efficiency of the method.
In the following we will use the simple interpolation given in \eq{gfparam}.

Finally we discuss what happens if the relative gauge fixing is extended to the
entire lattice and is not restricted to the points inside the blocks.
Links which are unchanged after the pure gauge
update would change through a global minimization.
This could introduce additional
noise and indeed this is the case for the full
Dirac operator but not for the Schur complement.
The global minimization slightly improves the behavior of the
interpolated fields in \eq{gfparam} but this effect is not large and the
danger to run into negative eigenvalues as discussed above increases.

\section{Volume dependence of the exact acceptance rate \label{s_exactacc}}

We simulate QCD with $\Nf=2$ flavors of mass-degenerate quarks.
The action for the gauge field is the Wilson plaquette gauge action 
\cite{Wilson:1974sk}
\be
S_g=\beta\Sw(U) = \frac{\beta}{6}\sum_p\Re\Tr\{1-U(p)\}\,, \label{eq:Sw}
\ee
where $p$ runs over all oriented plaquettes (i.e., each plaquette is counted
with two orientations).
For the fermions we use the plain Wilson--Dirac operator \cite{Wilson:1974sk}
(without clover term and without smearing) with bare quark mass $m_0$,
whose action on a quark field $\psi$ is given by
\bea
&&(\Dw(U)+m_0)\psi(x) = (4+m_0)\psi(x) - \nonumber\\
&&\sum_{\mu=0}^3\frac{1}{2}\{U(x,\mu)(1-\gamma_\mu)\psi(x+\hat{\mu}) +
U(x-\hat{\mu},\mu)^{\dagger}(1+\gamma_\mu)\psi(x-\hat{\mu})\}\,. \label{eq:Dw}
\eea
The hopping parameter is defined as $\kappa=1/(2m_0+8)$.
In this section we simulate at
parameters $\beta=5.6$ and $\kappa=0.15825$.
Theses values corresponds to a lattice spacing $a=0.0717(15)\fm$ 
\cite{DelDebbio:2006cn} and a pseudoscalar mass
$\mps\approx404\MeV$ \cite{DelDebbio:2007pz} (determined on a larger 
$32\times24^3$ lattice).
%
%%%%%%%%%%%%%%%%%%%%%%%%%%%%%%%%%%%%%%%%%%%%%%%%%%%%%%%%%%%%%%%%%%%%%%%%%%%%%%%
\begin{table}[t]
\begin{center}
\begin{tabular}{c|c|c|l|l|c}
$i$ & $n_i$ & \mc{3}{c}{actions} & \mc{1}{c}{$\Pacc$} \\
 & & $S^{(0)}=\Sw$ & \mc{1}{c}{$S^{(1)}=\Sw^{\rm HYP}$} & 
   \mc{1}{c}{$S^{(2)}=S_b$} & \mc{1}{c}{} \\ \hline
0 & 500 & $\beta_0^{(0)}=\phantom{-}5.6918$  & \mc{1}{c}{-} & 
   \mc{1}{c}{-} & \mc{1}{c}{-} \\
1 & 1  & $\beta_1^{(0)}=-0.0196$ & \mc{1}{c}{$\beta_1^{(1)}=\phantom{-}0.0963$}  & 
   \mc{1}{c}{-} &  \mc{1}{c}{95\%} \\
2 & 1  & $\beta_2^{(0)}=-0.1187$ & \mc{1}{c}{$\beta_2^{(1)}=-0.0614$} &
   \mc{1}{c}{$\beta_2^{(2)}=\phantom{-}1.644$} &  \mc{1}{c}{76\%} \\
3 & 1  & $\beta_3^{(0)}=\phantom{-}0.0465$ & \mc{1}{c}{$\beta_3^{(1)}=-0.0349$}       &
   \mc{1}{c}{$\beta_3^{(2)}=-0.644$} & \mc{1}{c}{varies}
\end{tabular}
\end{center}
\caption{Optimal parameters for the 4-step PSMS algorithm (representative set) for
  plain Wilson fermions at $\beta=5.6$ and $\kappa=0.15825$.}
\label{tab:pW5p6params}
\end{table}
%%%%%%%%%%%%%%%%%%%%%%%%%%%%%%%%%%%%%%%%%%%%%%%%%%%%%%%%%%%%%%%%%%%%%%%%%%%%%%%
%

We implement a 4-step PSMS algorithm based on a 
domain decomposition with block size $4^4$ and on a hierarchy
of three acceptance-rejection steps. 
Our code is based on the freely available software package DD-HMC by
M. L\"uscher \cite{soft:DDHMC}.
In order to enhance the acceptance rates
we introduce parameters as explained in Appendix \ref{s_appc}.

In the first step we update the active links in the $4^4$ blocks, which 
amount to a fraction of about 9.4\% of all links.
The gauge proposal consists of 500 iterations of two
Cabibbo-Marinari heat-bath \cite{Cabibbo:1982zn} sweeps (with reversed
sequence of gauge link updates and random choice of SU(2) subgroups)
at the shifted coupling $\beta_0^{(0)}=5.6918$.
The gauge proposal is then subjected to a first acceptance-rejection
step containing a plaquette action $\Sw^{\rm HYP}$ like in \eq{eq:Sw} but where
the plaquettes are constructed from HYP smeared links with the parameters of
\cite{Hasenfratz:2001hp} (one level of smearing).
We do one iteration of this step with 95\% acceptance.
The resulting proposal goes into a second acceptance-rejection step containing
the action $S_b=\sum_{b \in \mathcal{C}}2\ln(\det(D_b))$ of
the block determinants (one iteration with 76\% acceptance).
We emphasize that the first and second acceptance-rejection (or filter) steps
are done block-wise and can be therefore parallelized.
Finally the gauge proposal which passed through the first two filter steps
enters the global acceptance-rejection step with the Schur complement
of the $4^4$ block decomposition. This is a stochastic acceptance-rejection
step performed according to \eq{staccN} using the interpolation
with intermediate fields $U_i,\;i=0,\ldots,N$ in \eq{gfparam}.
The optimal parameters can be tuned following the prescription given
in \sect{s_tuning}. We note that they depend only mildly on the global lattice 
volume and a representative set is listed in \tab{tab:pW5p6params}.
%
%%%%%%%%%%%%%%%%%%%%%%%%%%%%%%%%%%%%%%%%%%%%%%%%%%%%%%%%%%%%%%%%%%%%%%%%%%%%%%%
\begin{figure}[t]
 \begin{center}
   \includegraphics*[angle=0,width=.48\textwidth]{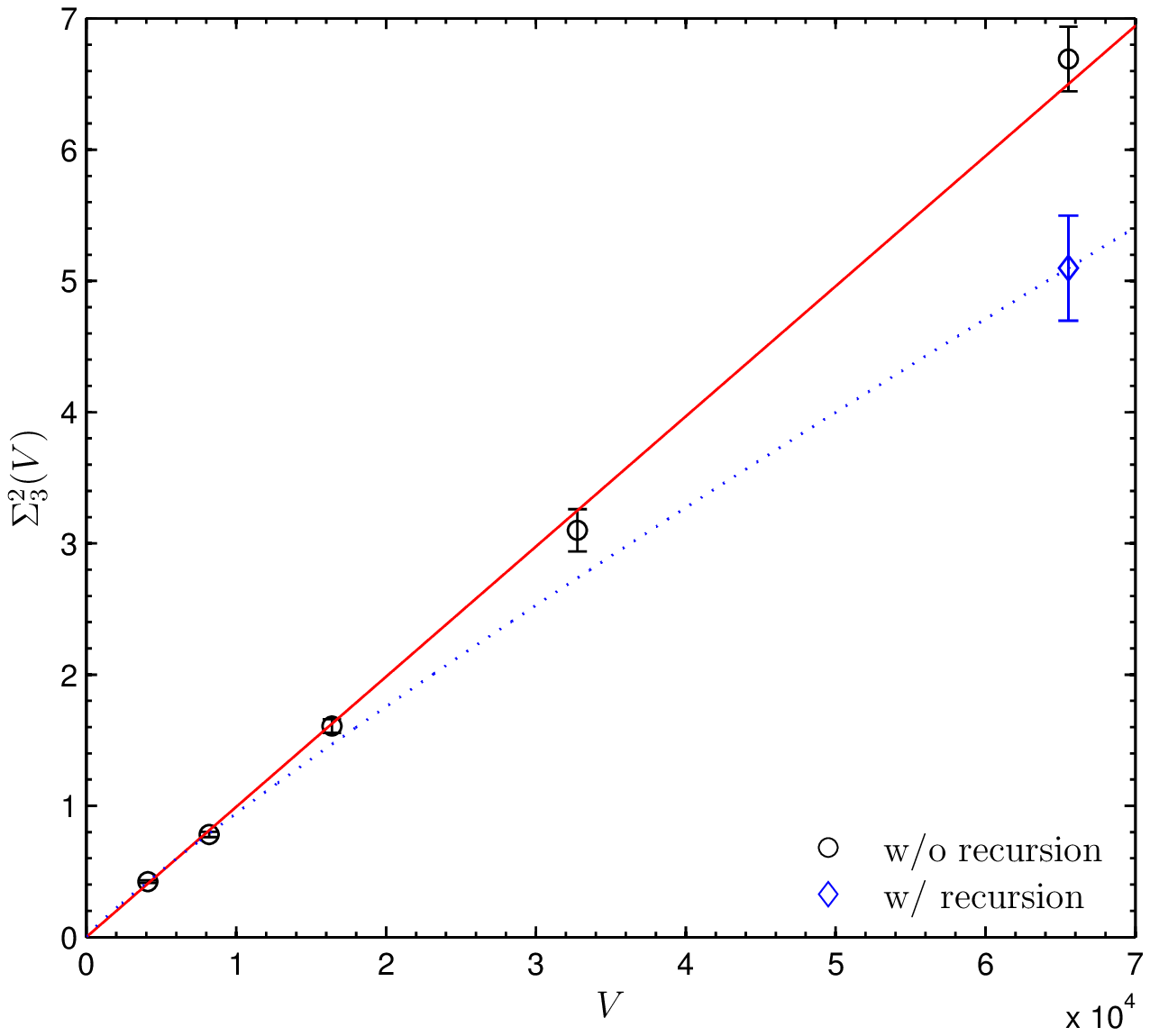}
   \includegraphics*[angle=0,width=.495\textwidth]{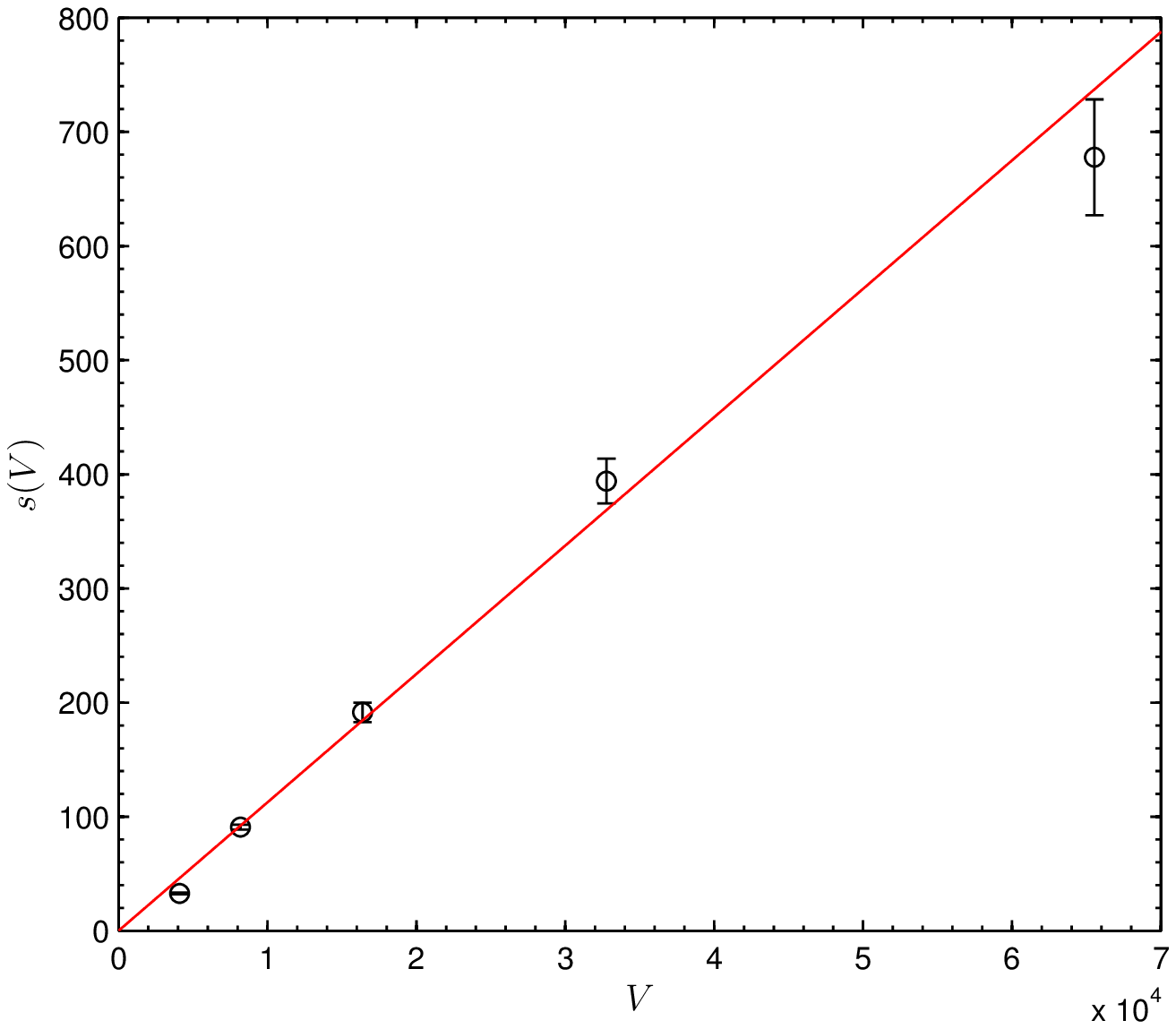}
 \end{center}
% \vspace{-0.5cm}
 \caption{\small
   The variance of the stochastic estimator in the global step
   from simulations of plain Wilson fermions at $\beta=5.6$, $\kappa=0.15825$
   with the 4-step PSMS algorithm.
   The left plot shows the exact variance $\Sigma^2_3$ (black circles)
   as a function of the lattice volume $V$ together with a linear fit (red line).
   The blue diamond is the result using a 5-step PSMS algorithm, see text.
   The right plot shows the volume dependence of the slope $s(V)$ defined
   in \eq{slopevar} together with a linear fit.
}
 \label{f_varglob}
\end{figure}
%%%%%%%%%%%%%%%%%%%%%%%%%%%%%%%%%%%%%%%%%%%%%%%%%%%%%%%%%%%%%%%%%%%%%%%%%%%%%%%
%

The global acceptance-rejection probability is
$\Pacc^{(3)} = \minm{\exp(-\Delta_3)}$, where (cf. \eq{DeltaSi}) 
\be
\Delta_3 = \beta_3^{(0)}\Delta \Sw+\beta_3^{(1)}\Delta \Sw^{\rm HYP}+
\beta_3^{(2)}\sum_{b \in \mathcal{C}}2\Delta \ln(\det(D_b))+
\sum_{i=0}^{N-1} \eta_i^\dagger (M_i^\dagger M_i-1) \eta_i
\ee
and $M_i$ is the ratio of Schur complements.
On lattices with $V=8^4$ up to $V=16^4$ we measure, for different
values of the gauge field interpolation steps $N$ in \eq{eq:param}, 
the variance
\be
\sigma_3^2(V,N)=\ev{(\Delta_3-\ev{\Delta_3})^2}_{U,\Up,\eta} \,.\label{varglob}
\ee
At fixed volume $V$ we extrapolate linearly in $1/N$ to zero,
thus obtaining an estimate for the exact variance $\Sigma^2_3(V)$
as a function of the volume.
The justification for this extrapolation is given by
\eq{varexactplusstoch}, which in this case means
\be
\sigma_3^2(V,N)=\Sigma^2_3(V)+
                \left(\sigma_{3,N}^{\rm stoch}(V)\right)^2 \label{var4step}
\ee
and by \eq{varstochapproxN}, which implies
\be
\left(\sigma_{3,N}^{\rm stoch}(V)\right)^2\approx \frac{1}{N}\,s(V) \,,
\label{slopevar}
\ee
with the slope $s(V)$ is approximately given by $\N1\,h_1^2$. Here
$\N1$ and $h_1$ refer to the Schur complement ratio.
Note that $\sigma_3^2(V,N)$ contains also contributions from parts of the
action other than the Schur complement (cf. \eq{sigmaDeltaSiparams}) but which
do not depend on $N$.
The extrapolated exact variance $\Sigma^2_3(V)$ is shown in the left plot 
of \fig{f_varglob} as a function of $V$.
The data can be very well fitted by a straight line
constrained to zero at zero volume (red line).
The slopes $s(V)$ of the linear fits of $\sigma_3^2(V,N)$ in $1/N$
are plotted against the volume $V$ in the right plot of \fig{f_varglob}. 
The data of the slope can be also well fitted by a straight line constrained
to zero at zero volume (red line).
Assuming that $\N1$ is equal to the dimension of the projector $P$
in \eq{dimproj} and taking into account that the block size is here constant
and equal to
$4^4$, we deduce that $\N1\propto V$. Therefore our results for the slope
means that the FWHM $h_1$ of the generalized eigenvalues
of the Schur complements does not significantly depend on the volume.
%
%%%%%%%%%%%%%%%%%%%%%%%%%%%%%%%%%%%%%%%%%%%%%%%%%%%%%%%%%%%%%%%%%%%%%%%%%%%%%%%
\begin{figure}[t]
 \begin{center}
     \includegraphics[width=0.60\textwidth]{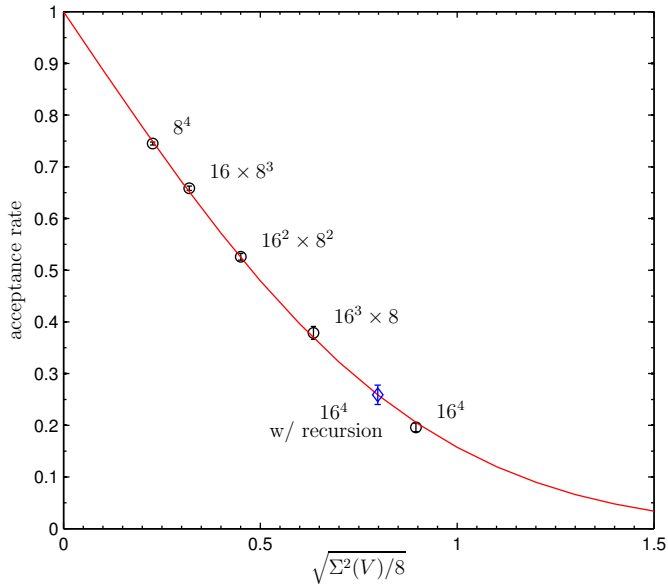}
 \end{center}
 \vspace{-0.5cm}
 \caption{\small The exact global acceptance is plotted as a function
   of the variance $\Sigma^2$ for
   simulations of plain Wilson fermions at $\beta=5.6$, $\kappa=0.15825$
   with the 4-step PSMS algorithm (black circles).
   The point corresponding to the blue diamond is from a simulation of a 
   $16^4$ lattice with a 5-step PSMS algorithm. 
}
 \label{f_exactacc}
\end{figure}
%%%%%%%%%%%%%%%%%%%%%%%%%%%%%%%%%%%%%%%%%%%%%%%%%%%%%%%%%%%%%%%%%%%%%%%%%%%%%%%
%

Via \eqref{eq:acceptance} the exact acceptance rate can be 
determined\footnote{We tested the (tacitly assumed)
validity of the Gaussian model for finite values of $N$.} from
the variance $\Sigma^2_3(V)$.
The exact acceptance rates as determined from the variances are plotted in
\fig{f_exactacc} (black circles)
together with the result from the fit to $\Sigma^2_3(V)$
shown in the left plot of \fig{f_varglob}. The 4-step PSMS
algorithm of this section shows a good acceptance for lattices up to
$16^3\times 8$.
This is the region where the error function can be approximated by a 
Taylor expansion with a linear leading term 
${\rm erfc}(x)=1-2x/\sqrt{\pi}+{\rm O}(x^3)$.
\fig{f_schuracc} shows that the acceptance rates, which one would obtain
from the Schur complement alone (blue diamonds), are much smaller.

The efficiency of the hierarchy of filters in enhancing the acceptance
of the global step
can be demonstrated by simulating the largest $16^4$ lattice using a
5-step PSMS algorithm. For this we use a recursive
domain decomposition of the $16^4$ lattice in $8^4$ and $4^4$ blocks,
cf. \eq{eq:ddrec}. The additional filter with respect to the 4-step PSMS
algorithm is a stochastic acceptance-rejection step accounting for the 
Schur complements of the $4^4$ blocks within the $8^4$ blocks, cf.~\eq{eq:4step}.
The acceptance of the global step (accounting for the global Schur complement)
is increased by this further filter step, cf.~the blue diamond in
 \fig{f_exactacc}.
Using recursive domain decomposition to keep the largest block size at
$L/2$ (where $V=L^4$), the volume dependence of $\Sigma^2$ in the global
step is $V^q$ (dotted line in the left plot of \fig{f_varglob})
with $q\approx0.9$ (determined on our available lattices $8^4$ and $16^4$).
At large $V$
one expects the asymptotic behavior $q=3/4$, cf.~\sect{subs_schur}.

\section{Numerical tests of the algorithm \label{s_res}}

We present results of simulations of $\Nf=2$ flavors of mass-degenerate
plain Wilson fermions on a $16^4$ lattice at $\beta=5.8$ and $\kappa=0.15462$. 
The clover coefficient is set to zero and the fermions have
anti-periodic boundary conditions in time direction.
The lattice spacing is estimated
in \cite{DelDebbio:2006cn} to be $0.0521(7)\,{\rm fm}$ and the
pseudoscalar mass is $381\,{\rm MeV}$ \cite{DelDebbio:2007pz} (determined
on a larger $64\times 32^3$ lattice).
%
%%%%%%%%%%%%%%%%%%%%%%%%%%%%%%%%%%%%%%%%%%%%%%%%%%%%%%%%%%%%%%%%%%%%%%%%%%%%%%%
\begin{table}[t]
\begin{center}
\begin{tabular}{c|c|c|c|l|l|c}
$i$ & $n_i$ & $N_i$ & \mc{3}{c}{actions} & \mc{1}{c}{$\Pacc$} \\
 & & & $S^{(0)}=\Sw$ & \mc{1}{c}{$S^{(1)}=\Sw^{\rm HYP}$} & 
   \mc{1}{c}{$S^{(2)}=S_b$} & \mc{1}{c}{} \\ \hline
0 & 75 & -  & $\beta_0^{(0)}=\phantom{-}5.9822$  & \mc{1}{c}{-} & 
   \mc{1}{c}{-} & \mc{1}{c}{-} \\
1 & 3  & -  & $\beta_1^{(0)}=-0.0378$ & \mc{1}{c}{$\beta_1^{(1)}=\phantom{-}0.2110$}  & 
   \mc{1}{c}{-} &  \mc{1}{c}{69\%} \\
2 & 3  & 96 & $\beta_2^{(0)}=-0.1376$ & \mc{1}{c}{$\beta_2^{(1)}=-0.2110$} &
   \mc{1}{c}{$\beta_2^{(2)}=\phantom{-}1.0711$} &  \mc{1}{c}{48\%} \\
3 & 1  & 96 & $\beta_3^{(0)}=-0.0068$ & \mc{1}{c}{$\beta_3^{(1)}=\phantom{-}0\phantom{.0000}$}       &
   \mc{1}{c}{$\beta_3^{(2)}=-0.0711$} & \mc{1}{c}{64\%}
\end{tabular}
\end{center}
\caption{Parameters for the 4-step PSMS algorithm for plain Wilson fermions
  at $\beta=5.8$ and $\kappa=0.15462$.}
\label{tab:pW5p8params}
\end{table}
%%%%%%%%%%%%%%%%%%%%%%%%%%%%%%%%%%%%%%%%%%%%%%%%%%%%%%%%%%%%%%%%%%%%%%%%%%%%%%%
%

In the simulations in \sect{s_exactacc} our smallest blocks are $4^4$ and
the gauge proposal changes the active links in these blocks. It turns out
that larger blocks are better in terms of changing the topological charge
and allow for higher global acceptances (at somewhat higher computational
cost). That is why we change our setup in this section and use $8^4$ blocks
as our smallest ones. The gauge proposal changes the active links in a
$6^4$ hypercube inside each of the $8^4$ blocks, which amounts to updating
$7.9$\% of all gauge links.

We adopt a 4-step PSMS algorithm whose parameters and acceptances
are summarized in \tab{tab:pW5p8params}. For each of the steps
$i=0,1,2,3$, $n_i$ is the number of iterations per step and $N_i$ is the
number of gauge field interpolation steps (for stochastic estimates of
Schur complement ratios).
The gauge proposal consists of a number $n_0$ of iterations of
symmetrized sweeps of Cabibbo-Marinari heat-bath
\cite{Cabibbo:1982zn} and over-relaxation \cite{Adler:1981sn,Petronzio:1990vx}
updates. One iteration consists of one heat-bath (HB) sweep
and $L/2$ over-relaxation (OR) sweeps followed by the reversed sequence of
link updates (so in total one HB + $L/2$ OR + $L/2$ OR + one HB sweeps),
where for
each link we choose with probability $1/2$ one sequence of SU(2)
subgroups and with probability $1/2$ the reversed sequence.
The first acceptance-rejection step is a Metropolis step for a HYP plaquette 
gauge action which has to be subtracted in the successive filter steps.
The determinant of the $8^4$ blocks is factorized by a domain decomposition in
$4^4$ blocks. The second acceptance-rejection step accounts for the exact
product of the $4^4$ block determinants times the determinant of the Schur
complement of the decomposition of the $8^4$ blocks in $4^4$ blocks. 
The latter is treated stochastically.
These two acceptance-rejection steps are performed independently for each 
$8^4$ block. The third stochastic acceptance-rejection step contains the global
Schur complement of the decomposition of the $16^4$ lattice in $8^4$ blocks. 
%
%%%%%%%%%%%%%%%%%%%%%%%%%%%%%%%%%%%%%%%%%%%%%%%%%%%%%%%%%%%%%%%%%%%%%%%%%%%%%%%
\begin{figure}[t]
  \begin{center}
   \includegraphics[angle=0,width=0.60\textwidth]{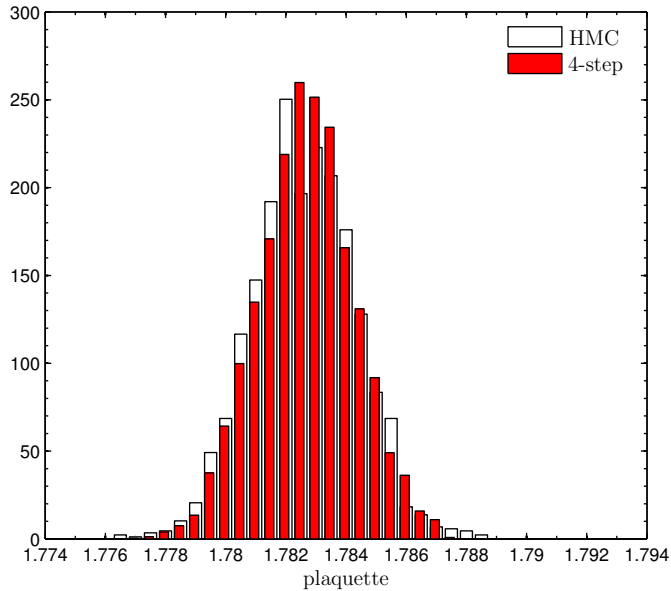}
\end{center}
  \caption{\small Histogram distribution of the plaquette values from
    simulations of plain Wilson fermions and
    Wilson plaquette action at $\beta=5.8$, $\kappa=0.154620$, $16^4$ lattices.
    We compare results for the 4-step PSMS algorithm and for the HMC.
}
  \label{fig:plaq_pW5p8}
\end{figure}
%%%%%%%%%%%%%%%%%%%%%%%%%%%%%%%%%%%%%%%%%%%%%%%%%%%%%%%%%%%%%%%%%%%%%%%%%%%%%%%
%

In \fig{fig:plaq_pW5p8} we show the histogram distribution of the plaquette
value. We compare the results from 4 replica simulated using the 4-step PSMS
algorithm (red bins) with the results from a long HMC simulation (white bins).
The HMC simulation is done with the DD-HMC algorithm 
\cite{Luscher:2005rx,Luscher:2007es} using $8^4$ blocks.
The distributions agree perfectly.
%
%%%%%%%%%%%%%%%%%%%%%%%%%%%%%%%%%%%%%%%%%%%%%%%%%%%%%%%%%%%%%%%%%%%%%%%%%%%%%%%
\begin{figure}[t]
 \begin{minipage}[c]{.48\textwidth}
  \begin{center}
   \includegraphics*[angle=0,width=\textwidth]{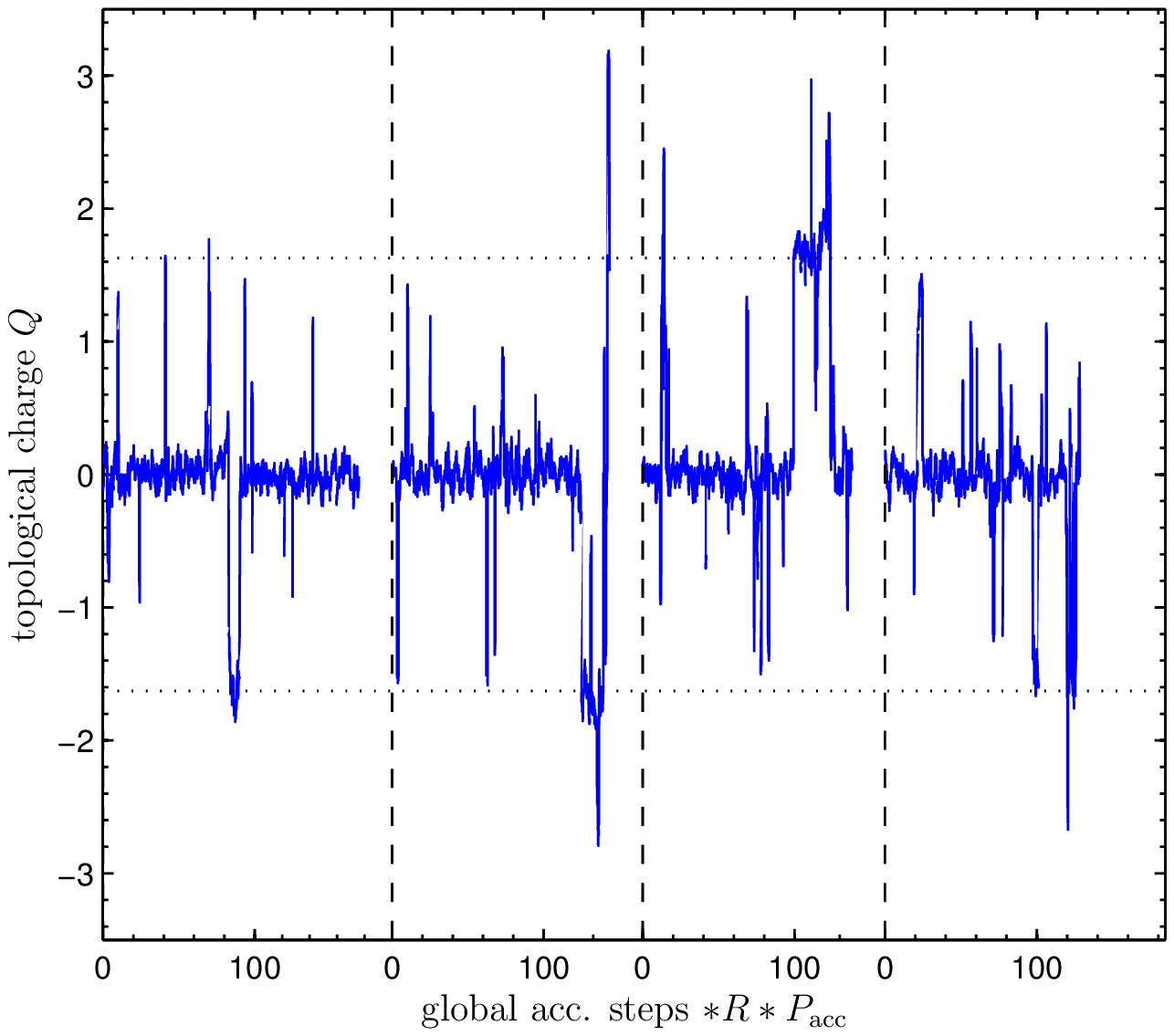}

   \vspace{0.5cm}
   \includegraphics*[angle=0,width=\textwidth]{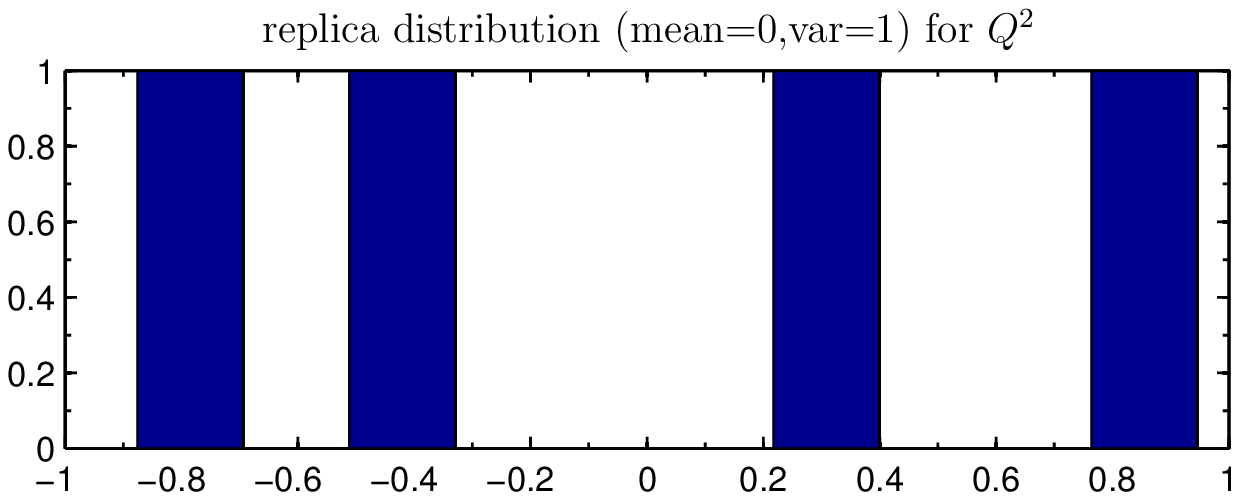}
   \end{center}
 \end{minipage}
 \hspace{.02\textwidth}
 \begin{minipage}[c]{.48\textwidth}
  \begin{center}
   \includegraphics*[angle=0,width=\textwidth]{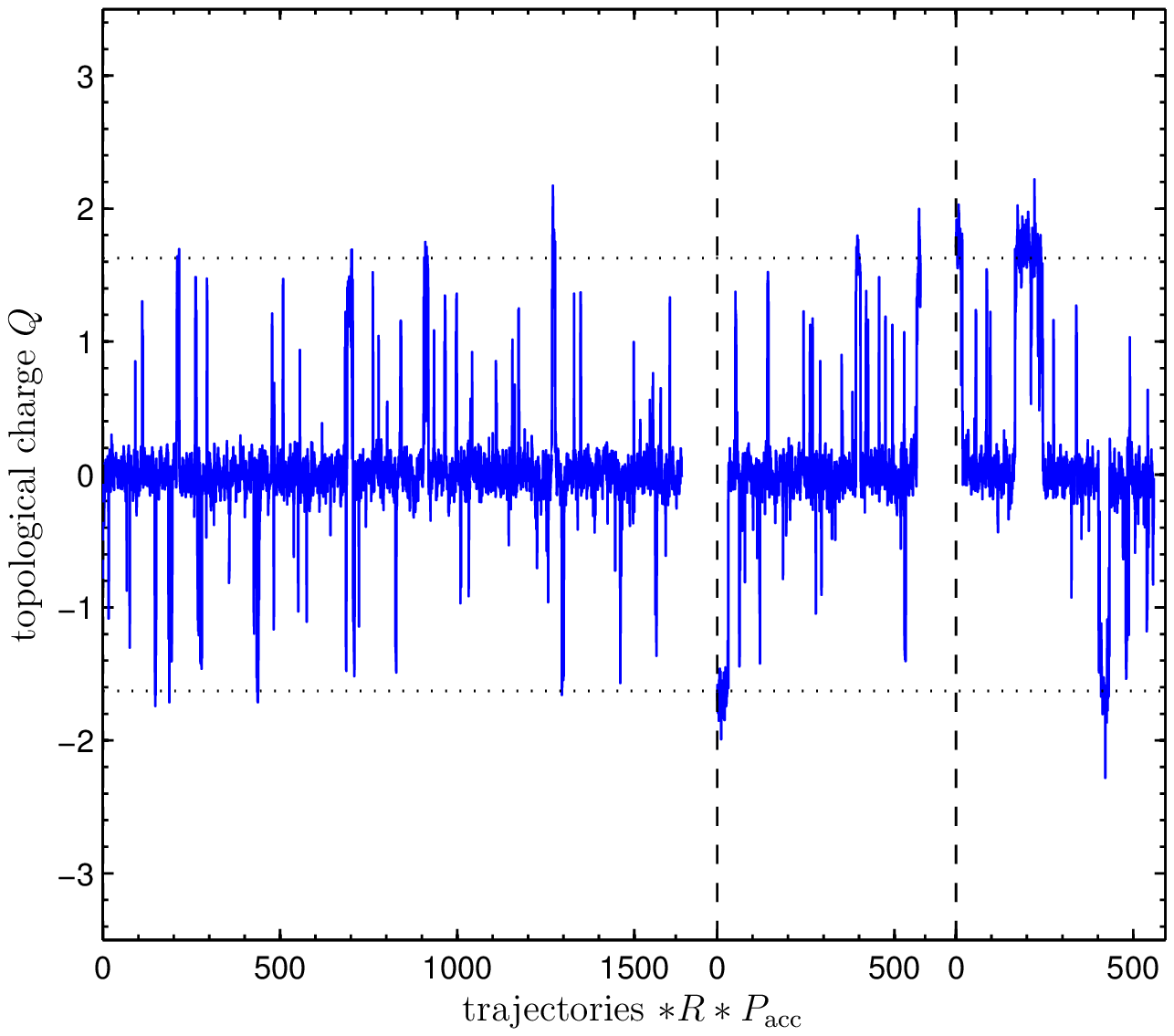}

   \vspace{0.5cm}
   \includegraphics*[angle=0,width=\textwidth]{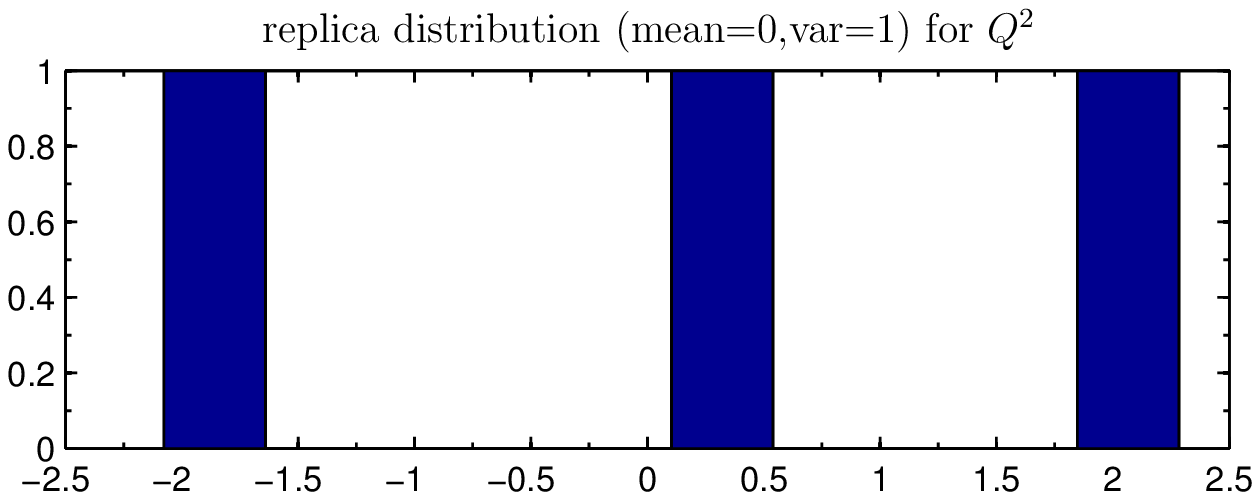}
  \end{center}
 \end{minipage}
% \vspace{-0.5cm}
 \caption{\small Histories of the topological charge $Q$ (upper plots) and
  histograms of the deviations of the replicum means of $Q^2$ from the total 
  mean divided by the replicum errors (lower plots).
  The left plots show the results of 4 replica simulated with the 4-step PSMS 
  algorithm. The right plots show the HMC results from 3 replica.
}
 \label{fig:topo_pW5p8}
\end{figure}
%%%%%%%%%%%%%%%%%%%%%%%%%%%%%%%%%%%%%%%%%%%%%%%%%%%%%%%%%%%%%%%%%%%%%%%%%%%%%%%
%

In the upper two plots of \fig{fig:topo_pW5p8} the histories of the topological 
charge are shown.
The topological charge is defined by
\be
Q = \frac{1}{16\pi^2}\sum_{x,\mu,\nu}F_{\mu\nu}(x)\tilde{F}_{\mu\nu}(x)\,,
\label{topoQ}
\ee
using a discretization of the field strength tensor $F_{\mu\nu}$ (see e.g.
\cite{Luscher:1996sc}) in which gauge links constructed from three levels of
HYP smearing are used.
We consider 4 replica simulated using the 4-step PSMS algorithm (left
plot) and 3 replica simulated with the HMC (right plot). The horizontal dotted
lines are determined from an ad hoc fit to the histogram of the topological charge using 3
Gaussian functions (one centered at zero and the other at values $\pm m$ corresponding
to the dotted lines). In order to compare the Monte Carlo histories of the two 
algorithms, we take the Monte Carlo units which correspond to a full change of the
gauge configuration. To this end, on the x-axis of the history plots we take,
for the PSMS algorithm, the number of global acceptance steps multiplied by the 
fraction $R$ of links changed and by the global acceptance while, for the HMC algorithm,
we take the number of trajectories multiplied by the ratio $R$ of active links and 
by the acceptance. The right plot shows that the long HMC replicum was not able to
really tunnel to a topological sector different than zero, while such a tunneling 
occurred at least once for all PSMS replica. Indeed we compared the distributions
of the topological charge squared $Q^2$ for the PSMS replica and the long HMC replicum
and found that they agree well around $Q^2=0$ but differ at larger values.
Therefore we started two more HMC replica from configurations with topological charge 
different than zero (generated in the PSMS ensembles), which are also shown in 
\fig{fig:topo_pW5p8}. In one of these two additional replica we observe a clear tunneling
from topological sector zero to nonzero. In the lower two plots of \fig{fig:topo_pW5p8}
we show histograms of the deviations of the replicum means of $Q^2$ from 
the total mean divided by the replicum errors 
(the quantity in Eq. (30) of \cite{Wolff:2003sm}; left plot, 
PSMS; right plot, HMC). The goodness of the replica distribution is measured by the
probability (goodness-of-fit) of a constant fit to the replicum means.
The goodness is 0.7 for the PSMS algorithm and 0.05 for the HMC. 
A value much below 0.1 is very unlikely. The expecation value $\ev{Q^2}$ is $0.37(15)$ 
for the 4-step PSMS algorithm and $0.281(81)$ for the HMC algorithm. The errors are 
determined using the method of \cite{Wolff:2003sm}. From leading order chiral perturbation 
theory we expect $\ev{Q^2}\approx0.19$. We emphasize that \fig{fig:topo_pW5p8} is a
comparison made at one lattice spacing only. The main problem is the scaling with
the lattice spacing which we cannot address in the scope of this paper.
%
%%%%%%%%%%%%%%%%%%%%%%%%%%%%%%%%%%%%%%%%%%%%%%%%%%%%%%%%%%%%%%%%%%%%%%%%%%%%%%%
\begin{figure}[t]
  \begin{center}
   \includegraphics*[angle=0,width=.48\textwidth]{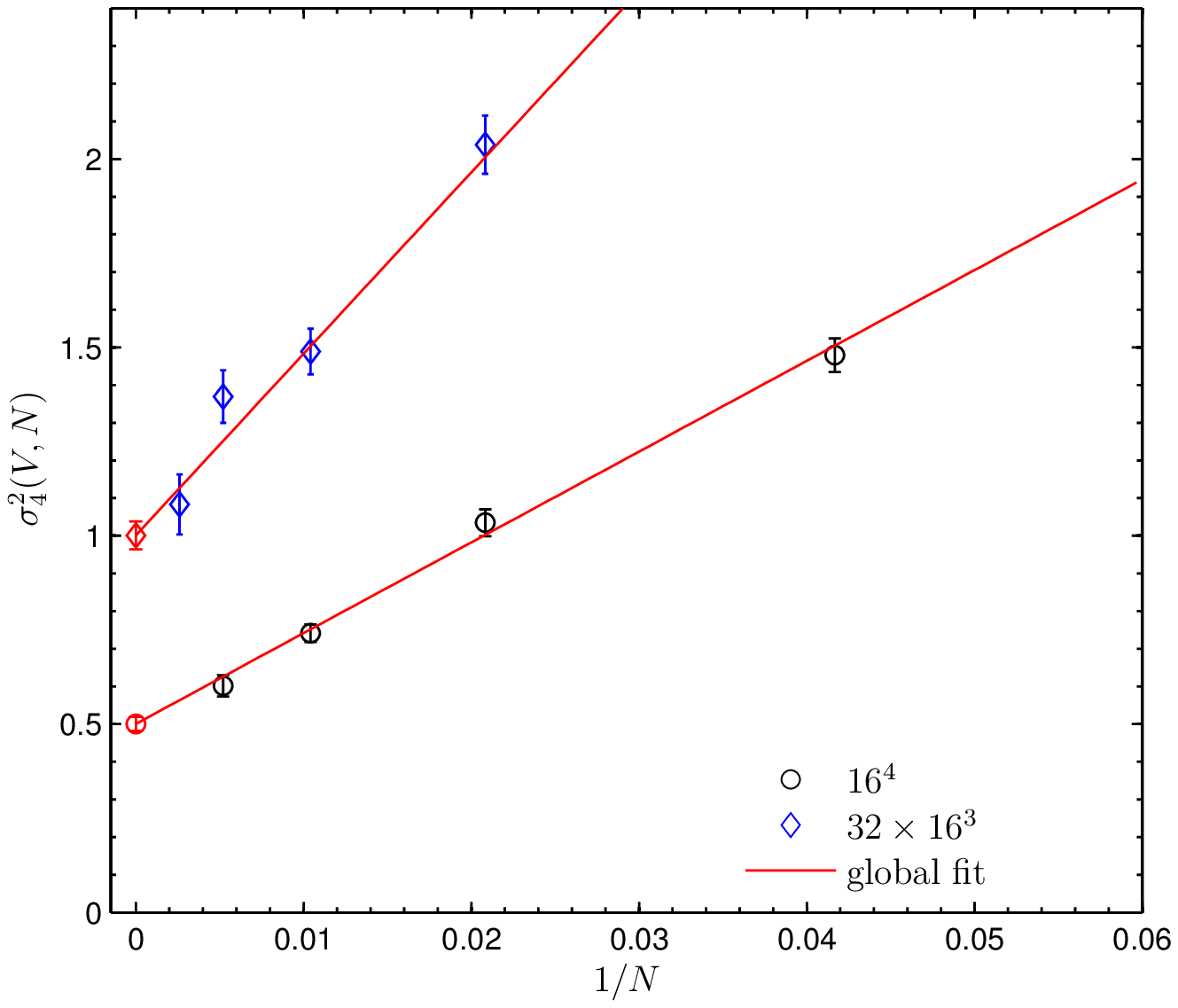}
   \includegraphics*[angle=0,width=.48\textwidth]{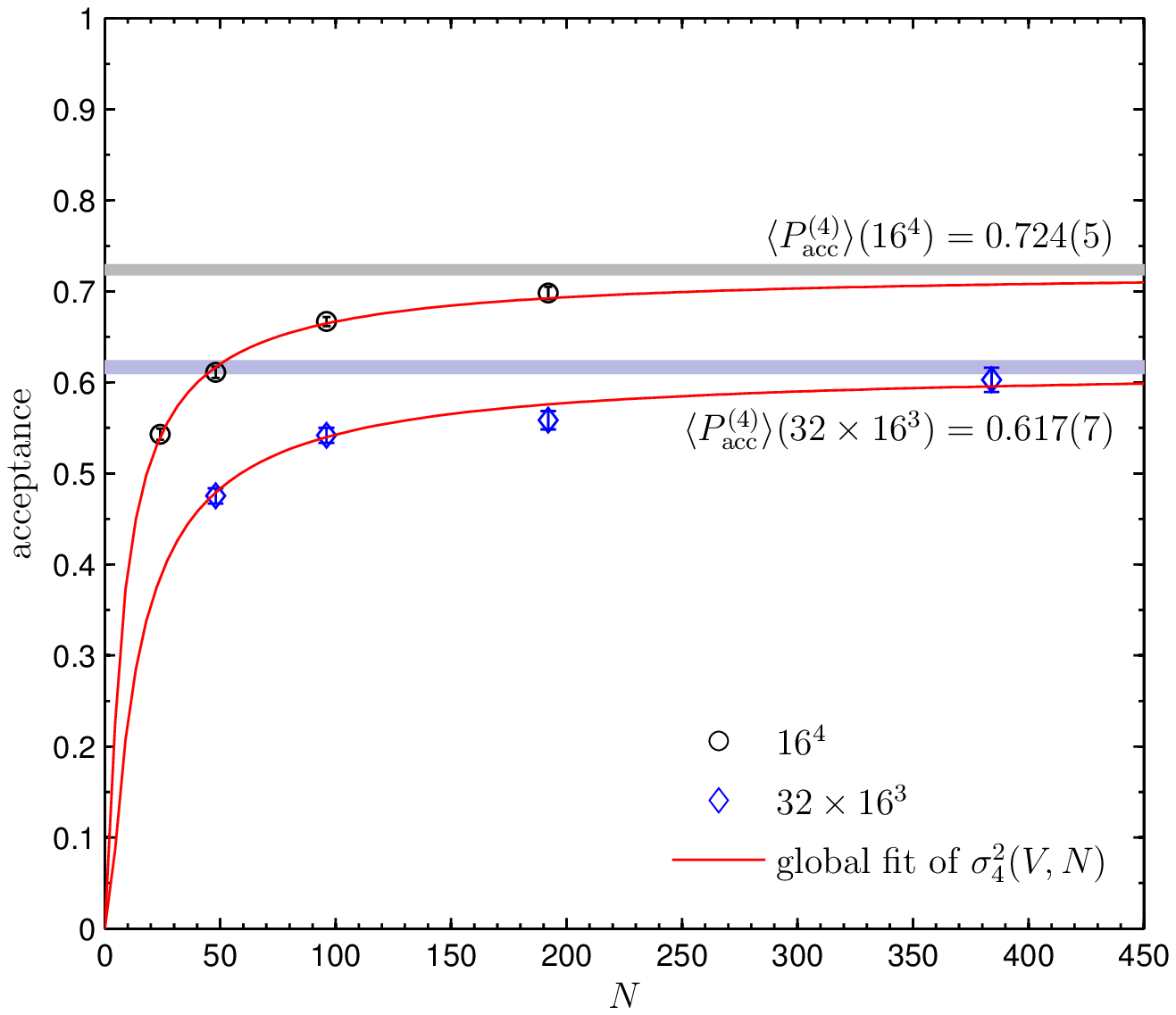}
\end{center}
  \caption{\small The variance $\sigma_3^2(V,N)$ (left plot) in the global
  step for simulations of plain Wilson fermions and
  Wilson plaquette action at $\beta=5.8$, $\kappa=0.154620$.
  Data for $V=16^4$ (circles) and $V=32\times16^3$ (diamonds) are very well 
  fitted as functions of $1/N$ using a global linear fit (red lines). In the
  right plot we show the resulting acceptances.
}
  \label{fig:sigma_pW5p8}
\end{figure}
%%%%%%%%%%%%%%%%%%%%%%%%%%%%%%%%%%%%%%%%%%%%%%%%%%%%%%%%%%%%%%%%%%%%%%%%%%%%%%%
%

In \fig{fig:sigma_pW5p8} we plot the variance $\sigma_3^2(V,N)$ (left plot)
and the acceptance (according to \eq{eq:acceptance}, right plot) in the
global acceptance-rejection step (see \eq{var4step}) as a function of
$1/N$ and $N$ respectively. Together with the data for the $16^4$ lattices we
present data
for $32\times16^3$ lattices. Motivated by the results of \sect{s_exactacc} we
perform a global fit to the variances of the form 
\be
\sigma_3^2(V,N) = V\,\left(a_1+\frac{a_2}{N}\right)
\ee
with fit parameters $a_1$ and $a_2$.
The exact variance turns out to be
$\Sigma^2_3=0.50(2)$ and $\Sigma^2_3=1.00(4)$ for the $16^4$ and 
$32\times16^3$ lattices respectively. This corresponds to acceptances $0.724(5)$
and $0.617(7)$.

A cost comparison of the simulations of the $16^4$ lattice
can be performed by comparing the
number of full inversions of the Wilson--Dirac operator needed to
update all the links.
Using the 4-step PSMS algorithm at optimal parameters, the global
acceptance-rejection step
has $N=96$ gauge field interpolation steps (each of which requires
one inversion of the full operator for the inversion of the global Schur 
complement, see \eq{inverseschur}) and 64\% acceptance. 
This means that we need $\approx23$ global steps or 2200 inversions 
to get a new gauge configuration.
If we instead run the 4-step PSMS algorithm with $N=24$ and
42\% global acceptance,
one new gauge configuration is obtained after $\approx35$ global steps or
840 inversions. The DD-HMC needs only 120 inversions for
one new gauge configuration.
This naive cost comparison does not take into account effects of autocorrelation
times, which are hard to estimate for observables like the topological charge.

\section{Conclusions \label{s_concl}}

We have developed and tested the PSMS algorithm for lattice QCD that consists of a
hierarchical filter of acceptance-rejection steps. The hierarchy is based on an
exact factorization of the fermion determinant. Although other factorization
are possible, we here deploy (recursive) domain decomposition as it separates
the determinant in a local (blocks) and global part (Schur complement).

We were able to determine the exact global acceptance rates for volumes up to 
$(1.2\; {\rm fm})^4$ and demonstrate that the filter is successful in
fighting the exponential decrease with the volume.

The global acceptance-rejection step with the Schur complement remains expensive. We estimate a
factor of ten in comparison with the HMC for the setup of Section \ref{s_res}. 
The expected scaling of the cost of the algorithm with the volume is
\[
V\;\mbox{(inversion)}\times V^{3/4}\;(N)\times 1/(\mbox{acceptance})\,.
\]
The first factor is due to the cost of one inversion of the Dirac operator and
the second factor arises from the necessity to keep the stochastic noise low.
A constant global acceptance is achieved for constant variance $\Sigma^2$ of
the action differences that go into the global step, i.e., $\varhl\propto 1/V$
is needed (cf. \eq{varexapprox}). Instead
we find $\varhl\sim{\rm const}$ as $V$ is increased (cf. \fig{f_varglob}).
Previously the fluctuations of the small eigenvalues of $\sqrt{D^\dagger D}$
have been found to decrease like $1/V$ \cite{Luscher:2008tw}. We do not seem
to see this behavior for $\varhl$.
The reason might be that our separation scale,
given by the inverse block size $1/L_{\rm b}$, is too large.
For the simulations at $\beta=5.8$ ($\beta=5.6$) with a block size of 8 
this scale is approximately $500\MeV$ ($360\MeV$).

At the moment the performance of the PSMS algorithm is worse than the one of the
HMC algorithm, but the scaling of autocorrelation times of the topological charge 
with the lattice spacing has to be studied to make a definite conclusion. 
In \fig{fig:topo_pW5p8} we present evidence that the PSMS algorithm is more
efficient in sampling the topological sectors compared to the HMC. It is
still relevant to study alternatives to the HMC and there are prospects of using
and improving the PSMS algorithm. One possibility is to apply reweighting for the 
Schur complement, cf. \cite{Finkenrath:2012cz} where we demonstrate that reweighting 
factors for the Schur complement have a better scaling with the volume compared to 
the full operator. Improved gauge actions can be included in the hierarchy of 
acceptance steps and there is room for better choices of the gauge updates within 
the blocks. Also factorizations of the determinant other than domain-decomposition 
could be used.

The techniques for the stochastic estimation of determinant ratios, which we 
introduced in this article for the acceptance-rejection steps, can be equally well 
applied to the case of reweighting, e.g., in the quark mass \cite{Finkenrath:2012cz}
or to account for electromagnetic effects.

\bigskip

{\bf Acknowledgement.} We thank Tony Kennedy for correspondence on the proof
of detailed balance, Martin L{\"u}scher for a clarification on the fluctuations
of small eigenvalues of the Dirac operator,
Rainer Sommer for discussions and Ulli Wolff for comments
on the relative gauge fixing. The Monte Carlo simulations were carried out 
on the cluster Stromboli at the University of Wuppertal and we thank the
University.

\begin{appendix}
\section{Proofs of detailed balance \label{s_appa}}

%%%%%%%%%%%%%%%%%%%%%%%%%%%%%%%%%%%%%%%%%%
\subsection{Exact acceptance-rejection steps \label{2stepex}}
%%%%%%%%%%%%%%%%%%%%%%%%%%%%%%%%%%%%%%%%%%

The simplest setup of our algorithm is to split up the gauge weight in
\eq{2flavordistr} from the
fermionic one. The idea is to propose a new gauge configuration $\Up$ by a
pure gauge updating algorithm and accept or reject it
by a Metropolis step accounting for the fermionic weight.
Let $T_0(U\rightarrow\Up)$ be the transition probability for the pure gauge
proposal which has to satisfy detailed balance for the distribution (see below)
\be
P_0(U) = \frac{\exp(-S_g(U))}{Z_0} \,, \label{distrib0}
\ee
where $Z_0$ is the partition function for the gauge action $S_g$.
The Metropolis acceptance-rejection step \cite{Metropolis:1953am} consists of
accepting or rejecting the proposal $\Up$ with probability
\be
\pacc(U,\Up) =
\min \left\{1,\frac{P_0(U)P(\Up)}{P(U)P_0(\Up)} \right\} \,.
\label{paccexact}
\ee
The transition probability for this algorithm is
\bea
&&T(U\rightarrow\Up) = T_0(U\rightarrow\Up)\pacc(U,\Up)
\nonumber \\
& & + \delta(U-\Up)\left(1-\int
D[\Upp]\, T_0(U\rightarrow\Upp)\pacc(U,\Upp)\right) \,. \label{transp}
\eea
In order for $T$ to satisfy detailed balance for the distribution $P$ in
\eq{2flavordistr}, $T_0$ has to satisfy detailed balance for the distribution $P_0$
in \eq{distrib0}.
If the gauge proposal is a sequence of gauge link updates, their order has to
be symmetrized or chosen randomly \cite{Hasenbusch:1998yb}.

%%%%%%%%%%%%%%%%%%%%%%%%%%%%%%%%%%%%%%%%%%%%%%%%%%%%%%%%%%%%%%%
\subsection{Stochastic acceptance-rejection steps \label{s_stoch}}
%%%%%%%%%%%%%%%%%%%%%%%%%%%%%%%%%%%%%%%%%%%%%%%%%%%%%%%%%%%%%%%

The exact calculation of the determinant ratio in \eq{paccexact} is
numerically prohibitive. It can be replaced by a stochastic approximation that
maintains detailed balance exactly \cite{Knechtli:2003yt}.

We follow closely Appendix A, in particular section A.5, of
\cite{Luscher:2003vf}. The variables of the system (the
gauge field) are enlarged by adding auxiliary stochastic variables, which are
called {\em pseudofermions} and are only used in the stochastic
acceptance-rejection step.
The equilibrium probability distribution for the enlarged system
of gauge field $U$ and pseudofermion $\eta$ is
\be
\hP(\eta,U) = \frac{{\rm e}^{-|D(U)^{-1}\eta|^2}\exp(-S_g(U))}{Z} \,.
\label{combined2stepstoch}
\ee
The pseudofermion is a complex-valued field $\eta$ with the measure
\be
D[\eta] = \prod_{x,\alpha}\frac{d\Re(\eta_{x,\alpha})
  d\Im(\eta_{x,\alpha})}{\pi} \,, \label{pfmeasure}
\ee
where the index $\alpha$ contains spin and color degrees of freedom.
The norm squared of $\eta$ is defined by the scalar product $(\eta,\eta)$:
\be
|\eta|^2 = (\eta,\eta) = \sum_{x,\alpha}\eta_{x,\alpha}^*\eta_{x,\alpha} \,.
\ee
The equilibrium distribution of the gauge field alone is recovered by
integrating over the pseudofermion:
\be
P(U) = \int \Deta \, \hP(\eta,U) \,. \label{average}
\ee
We also define the conditional probability
\be
\hP(\eta|U) = \frac{\hP(\eta,U)}{P(U)} = 
\frac{{\rm e}^{-|D(U)^{-1}\eta|^2}}{|\det D(U)|^2} \label{conditional2stepstoch}
\ee
to generate the pseudofermion field $\eta$ given the gauge field $U$.

The algorithm to update the enlarged system consists of alternating two Markov
steps.
The first is a global heatbath step for updating the
pseudofermion at given gauge field $U$. A new pseudofermion $\eta$ distributed
according to $\hP(\eta|U)$ in \eq{conditional2stepstoch} is generated through
\be
\eta = D(U) \xi \,, \label{heatbath}
\ee
where $\xi$ is a Gaussian random pseudofermion generated with probability
distribution $p(\xi)=\exp(-|\xi|^2)$ \footnote{
The pseudofermion measure in \eq{pfmeasure} is normalized such that $\int
D[\xi] p(\xi)=1$.}. The second step is a Metropolis step for the gauge field
at given pseudofermion. A new gauge field $\Up$ is proposed with
transition probability $T_0(U\to\Up)$, which satisfies detailed balance for
the distribution $P_0(U)$ in \eq{distrib0}. The proposal is followed
by an acceptance-rejection step with probability
\be
\min \left\{1,\frac{P_0(U)\hP(\eta,\Up)}{\hP(\eta,U)P_0(\Up)} \right\} =
\min\left\{1,\frac{{\rm e}^{-|D(U^\prime)|^2\eta}}{{\rm
    e}^{-|D(U)|^2\eta}}\right\} \,. \label{metropf}
\ee
Both the heatbath and Metropolis steps separately fulfill detailed balance
with respect to the combined probability distribution $\hP(\eta,U)$ in
\eq{combined2stepstoch} \cite{Hasenbusch:1998yb}. Therefore also their
composition has the correct fixed point probability \cite{Kennedy:2006ax}. 

We consider now a composite update step consisting of an heatbath update for
the pseudofermion in \eq{heatbath} immediately followed by a Metropolis step
for the gauge field in \eq{metropf}. If after this we forget the pseudofermion
field, this can be viewed as an update for the gauge field alone with
acceptance probability\footnote{
We thank Tony Kennedy for clarifying this point in a correspondence.}
\bea
\pacc(U,\Up) & = & \int \Deta \, \hP(\eta|U) \min \left\{ 1 ,
\frac{P_0(U) \hP(\eta,\Up)}{\hP(\eta,U) P_0(\Up)} \right\} \nonumber \\
& = & \int D[\xi]\,{\rm e}^{-|\xi|^2}\min\left\{1,{\rm
  e}^{-|M\xi|^2+|\xi|^2}\right\} \,, \label{paccstoch}
\eea
where the ratio operator $M$ is defined in \eq{Mdef}. 
The associated transition probability in \eq{transp}, where now $\pacc(U,\Up)$
is given by \eq{paccstoch}, satisfies detailed balance for the
equilibrium probability $P(U)$ due to the property \cite{Luscher:2003vf}
\be
[P(U)/P_0(U)]\,\pacc(U,\Up) = [P(\Up)/P_0(\Up)]\,\pacc(\Up,U)
\,, \label{detbalstoch1}
\ee
or equivalently \cite{Knechtli:2003yt}
\be
\frac{\pacc(U,\Up)}{\pacc(\Up,U)}=|\det(M)|^{-2} \,.
\label{detbalstoch2}
\ee
In practice, the acceptance step \eq{paccstoch} is done by drawing {\em one}
Gaussian distributed pseudofermion $\xi$ and accepting or rejecting depending
on the argument under the $\min$ function. We note that
it is not possible to perform the
average of the argument {\em under} the $\min$ function over many
pseudofermions, as this violates detailed balance.
The acceptance probability in \eq{paccstoch} was computed in
\cite{Knechtli:2003yt}
\be
\Pacc(U,\Up) =
\sum_{i} \; \min(1,1/\lambda_i)
\prod_{j\not=i} \; \frac{\lambda_i-1}{\lambda_i-\lambda_j}
\ee
in terms of the eigenvalues $\lambda_i$ of $M^\dagger M$. It is bounded
by the exact (non-stochastic) acceptance probability in \eq{paccexact}
\cite{Knechtli:2003yt}
\be
\pacc(U,\Up) \le \min\left\{1,|\det(M)|^{-2}\right\} \,.
\ee

So far we discussed the case of a proposal followed by an acceptance-rejection
steps. \eq{paccexact} can be generalized to an arbitrary number of
acceptance-rejection steps as discussed in \sect{s_hierarchy}. The algorithm
satisfies detailed balance and this is also true if (some of) the Metropolis
acceptance-rejections steps are replaced by their stochastic counterpart
\eq{paccstoch}.

%%%%%%%%%%%%%%%%%%%%%%%%%%%%%%%%%%%%%%%%%%%%%%%%%%%%%%%%%%%%%%%%%%%%%%%%%%%%%%%%
\subsection{Gauge field interpolation \label{s_stochpar}}
%%%%%%%%%%%%%%%%%%%%%%%%%%%%%%%%%%%%%%%%%%%%%%%%%%%%%%%%%%%%%%%%%%%%%%%%%%%%%%%%

In order to simplify a bit the notation we consider an algorithm like it is
described in \sect{s_stoch} with one stochastic acceptance-rejection step. In
practice we apply the gauge field interpolation method to
acceptance-rejection steps involving the Schur complements (the global Schur
complement $\hat{D}$ as well as the Schur complement in the blocks $\hat{D}_b$
when we use recursive domain decomposition, see \eq{eq:ddrec}).

For the gauge proposal $U\to\Up$ we consider the gauge field interpolation
$U_i$ as it is given in \eq{gfparam}.
For each of the transitions $U_i\to U_{i+1}$, $i=0,1,\cdots,N-1$ we introduce
a pseudofermion field $\eta_i$. 
The equilibrium probability distribution for the enlarged system is
\be
\hat{P}(\{\eta_j\},U,\Up)=
\frac{{\rm e}^{-|D(U^{g})^{-1}\eta_0|^2} {\rm e}^{-S_g(U)}}{Z}
\prod_{i=1}^{N-1} \frac{{\rm e}^{-|D(U_i)^{-1}\eta_i|^2}}{|\det(D(U_i))|^2}
\ee
and depends now also on the proposed configuration $\Up$.
Integrating over the pseudofermions gives
\be
P(U) = \int\prod_{i=0}^{N-1} D[\eta_i] \, \hat{P}(\{\eta_j\},U,\Up) \,.
\ee
The conditional probability to generate the pseudofermions $\{\eta_j\}$ given
the proposal $U\to\Up$ is
\be
\hP(\{\eta_j\}|U,\Up) = \frac{\hat{P}(\{\eta_j\},U,\Up)}{P(U)} =
\prod_{i=0}^{N-1}
\frac{{\rm e}^{-|D(U_i)^{-1}\eta_{i}|^2}}{|\det(D(U_i))|^2} \,.
\ee
We use the property $\det(D(U))=\det(D(U^{g}))$.

If we consider the reversed gauge proposal $\Up\to U$ (i.e. $U_0={\Up}^{g^{-1}}$
and $U_N=U^{g}$), the intermediate
configurations $U_i$ in \eq{gfparam} are the same but they are traversed in
reversed order and therefore the pseudofermion $\eta_i$ is associated with the
transition $U_{i+1}\to U_i$. 
The probability distribution for the enlarged system is now
\be
\hP(\{\eta_j\},\Up,U)=
\frac{{\rm e}^{-|D({\Up}^{g^{-1}})^{-1}\eta_{N-1}|^2} {\rm e}^{-S_g(\Up)}}{Z}
\prod_{i=0}^{N-2} \frac{{\rm e}^{-|D(U_{i+1})^{-1}\eta_i|^2}}{|\det(D(U_{i+1}))|^2}
\ee
and the conditional probability to generate the pseudofermions $\{\eta_j\}$ is
\be
\hP(\{\eta_j\}|\Up,U) = \frac{\hat{P}(\{\eta_j\},\Up,U)}{P(\Up)} =
\prod_{i=0}^{N-1}
\frac{{\rm e}^{-|D(U_{i+1})^{-1}\eta_{i}|^2}}{|\det(D(U_{i+1}))|^2} \,.
\ee

The acceptance probability for the gauge proposal $U\to\Up$ is
\bea
\pacc(U,\Up)
&=&\int \prod_{i=0}^{N-1} D[\eta_i] \, \hP(\{\eta_j\}|U,\Up) 
\min\left\{1,\frac{P_0(U) \hP(\{\eta_j\},\Up,U)}
                  {\hP(\{\eta_j\},U,\Up) P_0(\Up)}\right\} \nonumber \\
&=& \int \prod_{i=0}^{N-1} D[\xi_i] \, {\rm e}^{-|\xi_i|^2}
\min\left\{1,{\rm e}^{\sum_{j=0}^{N-1}-|M_j\xi_j|^2+|\xi_j|^2}
\right\} \,,\label{paccstochgp}
\eea
where 
\be
M_i = D(U_{i+1})^{-1} D(U_i) \,.
\ee
$\pacc(U,\Up)$ in \eq{paccstochgp}
fulfills the detailed balance condition \eq{detbalstoch1} or equivalently
\be
\frac{\pacc(U,\Up)}{\pacc(\Up,U)} = |\det(M)|^{-2} \,,\quad
M=D(\Up)^{-1}D(U)\,.
\ee
In practice, the global acceptance step \eq{paccstochgp} is done by drawing
$N$ Gaussian distributed pseudofermions $\xi_i$ and accepting or rejecting
depending on the argument of the $\min$ function (i.e. we evaluate the sum in
the exponent under the $\min$ function).

\section{Relative gauge fixing \label{s_appb}}

The relative gauge fixing of two gauge field configurations $U$ and $\Up$ 
is done using a steepest descent scheme introduced by
\cite{Batrouni:1985jn,Davies:1987vs} for gauge group SU(3). 
Using the condition \eq{choiceg} for the gauge transformation $g$,
the minimization condition in \eq{relgf}
can be written similarly to the case of the Landau gauge condition. Then one
can apply the procedure of \cite{Davies:1987vs} to fix the relative gauge.

At each point $x$ where the gauge transformation $g$ is defined we
have to solve the condition
\be
\min_{g(x)} \Re\Tr \left\{1-
(g(x)^\dagger)^2\cdot (W_{\rm f}(x) + W_{\rm b}(x) \right\} \,,
\label{eq:GF:fminendform}
\ee
where
\bea
W_{\rm f}(x) & = & \sum_{\mu} 
\Up(x,\mu)g(x+\hat{\mu})^2U^\dagger(x,\mu) \,,\\
W_{\rm b}(x) & = & \sum_{\mu} 
{\Up}^\dagger(x-\hat{\mu},\mu)g(x-\hat{\mu})^2U(x-\hat{\mu},\mu) \,. 
\eea
Using the steepest descent method of \cite{Davies:1987vs} we get the
minimizing transformation field $g(x)$ iteratively through
\be
g(x)= \exp\left\{ -\frac{\alpha}{2}\left[
\Delta - \Delta^\dagger -\frac{1}{3} \Tr(\Delta - \Delta^\dagger)
\right]\right\} \label{ming}
\ee
with a scaling parameter $\alpha$ and 
\be
\Delta = W_{\rm f}(x) + W_{\rm b}(x) \,.
\label{eq:GF:Delta}
\ee
The minimum is reached when $\theta(x)=0$, where
\be
\theta(x) = \Tr\left[
\Delta - \Delta^\dagger -\frac{1}{3} \Tr(\Delta - \Delta^\dagger) \right]^2
\,. \label{eq:GF:theta}
\ee
We choose the value $\alpha=0.15$. If there is no convergence we
reduce it in steps of $-0.01$ and reach values down to $\alpha=0.10$.
For the SU(3) exponential function in \eq{ming}
we use the matrix function described in Appendix A of \cite{Luscher:2005rx}.
The numerical cost of the relative gauge fixing can
be reduced in the case of a domain decomposition by defining $g(x)$
only inside the blocks where the active links are changed. 
Further it can be reduced by stopping the iteration when
$\theta(x)<10^{-3}$, which we find good enough for the purpose of the
gauge field interpolation discussed in \sect{subs_gfp}.

\section{Parametrized acceptance-rejection steps \label{s_appc}}
%
%%%%%%%%%%%%%%%%%%%%%%%%%%%%%%%%%%%%%%%%%%%%%%%%%%%%%%%%%%%%%%%%%%%%%%%%%%%%%%%
\begin{figure}[t]
 \begin{center}
     \includegraphics[width=0.60\textwidth]{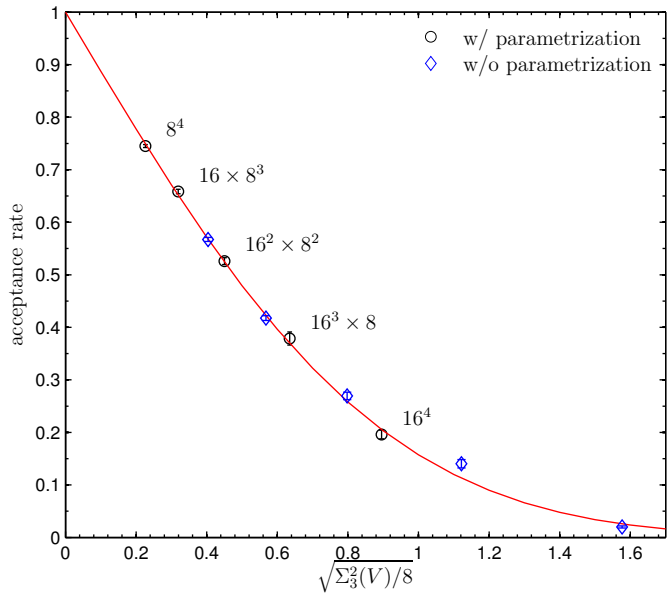}
 \end{center}
 \vspace{-0.5cm}
 \caption{\small The exact global acceptance is plotted as a function
   of the variance $\Sigma_3^2$ for simulations of plain Wilson fermions at 
   $\beta=5.6$, $\kappa=0.15825$ with the 4-step PSMS algorithm.
   The black circles are the same as in \fig{f_exactacc} and corresponds
   to the optimal acceptance. The blue diamonds represent the exact
   acceptance which one would get from the Schur complement alone without 
   the additional parameters (last row in \tab{tab:pW5p6params}).
}
 \label{f_schuracc}
\end{figure}
%%%%%%%%%%%%%%%%%%%%%%%%%%%%%%%%%%%%%%%%%%%%%%%%%%%%%%%%%%%%%%%%%%%%%%%%%%%%%%%
%

The general idea of the PSMS algorithm is to factorize the distribution in
\eq{2flavordistr} in several pieces and introduce a recursive update procedure with
a computational cost ordering.
Naively speaking a gauge configuration is proposed by a pure gauge algorithm
and the fermion determinant is treated in acceptance-rejection steps. 
It is easy to see that the plaquette gauge action and 
the determinant of the Dirac operator are strongly correlated. This correlation
can be used to increase the acceptance
\cite{Irving:1996bm,Hasenfratz:2002jn,Knechtli:2003yt} and also in the case of
reweighting \cite{Hasenfratz:2008fg}. 
This is an example of ultraviolet filtering
\cite{Hasenbusch:1998yb,deForcrand:1998sv,Duncan:1999xh}.

\subsection{Optimization of the acceptance \label{s_optacc}}

We consider the hierarchy of acceptance-rejection steps $i=1,2,\ldots,n$ in
\eq{eq:hierarchy}. The factorization \eqref{eq:factor} is not unique and
can be parametrized. In each acceptance-rejection step the action involved
might be written as
\be
S_i = \sum_{j=0}^i\beta_i^{(j)}S^{(j)} \,,\quad i=1,\ldots,n \,, \label{Si}
\ee
with real coefficients $\beta_i^{(j)}$ and actions $S^{(j)}$.
The probability to accept the proposal for a new gauge field
$\Up$ starting from $U$ is $\minm{\exp(-\Delta_i)}$ where
\be
\Delta_i = S_i(\Up)-S_i(U) = \sum_{j=0}^i\beta_i^{(j)}\Delta S^{(j)}\,. 
\label{DeltaSi}
\ee
For Gaussian distributed $\Delta_i$ the acceptance rate is given by
${\rm erfc}\left(\sqrt{\Sigma_i^2/8}\right)$ with
\be
\Sigma_i^2 = \ev{(\Delta_i-\ev{\Delta_i})^2} \,. \label{sigmaDeltaSi}
\ee
In the case of the factorization \eq{eq:ddfactor}, $n=2$ and 
$S^{(0)}$ is the gauge action,
$S^{(1)} = \sum_{b \in \mathcal{C}}2\ln(\det(D_b))$ and
$S^{(2)}=2\ln(\det(\hat{D}))$ are the effective actions of the
determinants of the blocks and of the Schur complement respectively.
In stochastic acceptance-rejection steps, like we do
for the Schur complement, we use
\be
\Delta S^{(2)} = \eta^\dagger (M^\dagger M-1) \eta \,,\label{Deltaistoch}
\ee
where $\eta$ is a Gaussian noise vector and $M=\hat{D}(\Up)^{-1}\hat{D}(U)$.
In such case
the variance $\Sigma_i^2$ in \eq{sigmaDeltaSi} is replaced
by the sum of the exact variance and the stochastic variance
according to \eq{varexactplusstoch}.
The variance in \eq{sigmaDeltaSi} can be written explicitly in terms of the
coefficients as
\be
\Sigma_i^2 = \sum_{j=0}^i\left(\beta_i^{(j)}\right)^2C^{(jj)}+
\sum_{\substack{j,k=0\\j\neq k}}^{i}\beta_i^{(j)}\beta_i^{(k)}C^{(jk)} \,,
\label{sigmaDeltaSiparams}
\ee
where $C^{(jk)}=\ev{(\Delta S^{(j)}-\ev{\Delta S^{(j)}}) (\Delta
  S^{(k)}-\ev{\Delta S^{(k)}})}$. We note that here $\ev{\cdot}$ means an
average over configurations $U$ in the dynamical ensemble and over gauge
proposals $\Up$ (and over noise $\eta$ if applicable).

At each step $i=1,\ldots,n$
the optimization of the parameters $\beta_i^{(j)}$ is done by
minimizing the variance $\Sigma_i^2$ in \eq{sigmaDeltaSi}.
The idea is to use the correlation of low cost actions with the high cost 
action of the $i$th step to increase the $i$th-level acceptance rate. 
In order to get the right distribution after the last step the parameters 
of a specific action $S^{(j)}$ has to sum up to the target value
$\beta^{(j)}=\sum_i\beta_i^{(j)}$. This implies constraints on the parameter.
For example the parameter of the plaquette action has to sum up to 
$\beta^{(0)} = \beta$. 
In principle, in order to solve for the parameters, we can start from the last
step $i=n$, solve for the parameters
$\beta_n^{(j)}$ and go to step $i-1$. This provides an explicit solution
scheme. At step $i$, 
we solve the linear system of $i$ equations
\be
2C^{(jj)}\beta_i^{(j)} + \sum_{\substack{k=0\\k\neq j}}^{i-1}C^{(jk)}\beta_i^{(k)}=
-C^{(ji)}\beta_i^{(i)}\,,\quad j=0,\ldots,i-1\,, \label{eq:paramsi}
\ee
to uniquely determine the values of the coefficients $\beta_i^{(0)},\ldots,
\beta_i^{(i-1)}$.
The solutions of the steps $k>i$ and one constraint imply
$\beta_i^{(i)}=\beta^{(i)}-\sum_{k=i+1}^n \beta_k^{(i)}$.

We emphasize some properties of the parametrized acceptance-rejection steps.
First of all this quite simple technique guarantees that the distribution of 
the pure gauge proposal has a good overlap with the dynamical distribution.
Without parametrization the acceptance rate for lattices bigger than $4^4$
would be less than few \%.
The parametrization introduces a $\beta$-shift to higher $\beta$ values in
the pure gauge update, mainly reflecting the correlation with the determinants
on the small blocks, see \tab{tab:pW5p6params} and \tab{tab:pW5p8params}.
In general it is possible to introduce a new
acceptance-rejection step $i$ by defining an auxiliary action with additional
 parameters.
These parameters have to sum up to zero when considering all
acceptance-rejection steps $k\ge i$. Their effect is to enhance the acceptance
rate of
these steps. For example we introduced a plaquette action, which uses HYP
smeared links (one level of smearing) in order to better match the pure gauge
update with the fermionic weight. 
This is particularly motivated for simulations with HYP smeared Wilson 
fermions but also helps for plain Wilson fermions, see \tab{tab:pW5p6params} 
and \tab{tab:pW5p8params}.
We remark that it is not possible to introduce parameters for
terms which are evaluated stochastically like \eq{Deltaistoch}.
The effectiveness of the parametrization of acceptance-rejection steps is
demonstrated in \fig{f_schuracc}, where we compare the exact acceptances
in the global step with optimal parameters (black circles) to the exact
acceptances without parameters (blue diamonds).

\subsection{Tuning the optimal parameters \label{s_tuning}}

The parameters in the acceptance-rejection steps $i=1,2,\ldots,n-1$ are
estimated from a simulation where the global step $i=n$ (the computationally
most costly) is left out.
Subsequently a full simulation is performed in order to determine the
optimal parameters for the global step. Iterating further this procedure
does not significantly change the values of the parameters.

\end{appendix}

\bibliography{algo}              %or whatever your .bib file is

\begin{thebibliography}{10}

\bibitem{Duane:1987de}
S. Duane, A. Kennedy, B. Pendleton and D. Roweth,
\newblock Phys.Lett. B195 (1987) 216.

\bibitem{Gottlieb:1987mq}
S.A. Gottlieb, W. Liu, D. Toussaint, R. Renken and R. Sugar,
\newblock Phys.Rev. D35 (1987) 2531.

\bibitem{Schaefer:2010hu}
ALPHA Collaboration, S. Schaefer, R. Sommer and F. Virotta,
\newblock Nucl.Phys. B845 (2011) 93, 1009.5228.

\bibitem{DelDebbio:2002xa}
L. Del~Debbio, H. Panagopoulos and E. Vicari,
\newblock JHEP 0208 (2002) 044, hep-th/0204125.

\bibitem{Luscher:2010iy}
M. L{\"u}scher,
\newblock JHEP 1008 (2010) 071, 1006.4518.

\bibitem{Luscher:2011kk}
M. L{\"u}scher and S. Schaefer,
\newblock JHEP 1107 (2011) 036, 1105.4749.

\bibitem{DelDebbio:2005qa}
L. Del~Debbio, L. Giusti, M. L{\"u}scher, R. Petronzio and N. Tantalo,
\newblock JHEP 0602 (2006) 011, hep-lat/0512021.

\bibitem{Durr:2010aw}
S. D{\"u}rr et~al.,
\newblock JHEP 1108 (2011) 148, 1011.2711.

\bibitem{Hasenfratz:2007rf}
A. Hasenfratz, R. Hoffmann and S. Schaefer,
\newblock JHEP 0705 (2007) 029, hep-lat/0702028.

\bibitem{Morningstar:2003gk}
C. Morningstar and M.J. Peardon,
\newblock Phys.Rev. D69 (2004) 054501, hep-lat/0311018.

\bibitem{Capitani:2006ni}
S. Capitani, S. D{\"u}rr and C. Hoelbling,
\newblock JHEP 0611 (2006) 028, hep-lat/0607006.

\bibitem{Kamleh:2004xk}
W. Kamleh, D.B. Leinweber and A.G. Williams,
\newblock Phys.Rev. D70 (2004) 014502, hep-lat/0403019.

\bibitem{Knechtli:2003yt}
Alpha Collaboration, F. Knechtli and U. Wolff,
\newblock Nucl.Phys. B663 (2003) 3, hep-lat/0303001.

\bibitem{Hasenfratz:2002jn}
A. Hasenfratz and F. Knechtli,
\newblock Comput.Phys.Commun. 148 (2002) 81, hep-lat/0203010.

\bibitem{Hasenfratz:2002pt}
A. Hasenfratz and A. Alexandru,
\newblock Nucl.Phys.Proc.Suppl. 119 (2003) 994, hep-lat/0209071.

\bibitem{Hasenfratz:2002vv}
A. Hasenfratz,
\newblock Nucl.Phys.Proc.Suppl. 119 (2003) 131, hep-lat/0211007.

\bibitem{Hasenfratz:2005tt}
A. Hasenfratz, P. Hasenfratz and F. Niedermayer,
\newblock Phys.Rev. D72 (2005) 114508, hep-lat/0506024.

\bibitem{Joo:2001bz}
B. Joo, I. Horvath and K. Liu,
\newblock Phys.Rev. D67 (2003) 074505, hep-lat/0112033.

\bibitem{Finkenrath:2011py}
J. Finkenrath, F. Knechtli and B. Leder,
\newblock (2011), 1112.1243.

\bibitem{Hasenbusch:1998yb}
M. Hasenbusch,
\newblock Phys.Rev. D59 (1999) 054505, hep-lat/9807031.

\bibitem{Luscher:2005rx}
M. L{\"u}scher,
\newblock Comput.Phys.Commun. 165 (2005) 199, hep-lat/0409106.

\bibitem{Luscher:2003vf}
M. L{\"u}scher,
\newblock JHEP 0305 (2003) 052, hep-lat/0304007.

\bibitem{Metropolis:1953am}
N. Metropolis, A. Rosenbluth, M. Rosenbluth, A. Teller and E. Teller,
\newblock J.Chem.Phys. 21 (1953) 1087.

\bibitem{Irving:1996bm}
A.C. Irving and J.C. Sexton,
\newblock Phys.Rev. D55 (1997) 5456, hep-lat/9608145.

\bibitem{Weingarten:1980hx}
D. Weingarten and D. Petcher,
\newblock Phys.Lett. B99 (1981) 333.

\bibitem{Luscher:2008tw}
M. L{\"u}scher and F. Palombi,
\newblock PoS LATTICE2008 (2008) 049, 0810.0946.

\bibitem{Hasenbusch:2001ne}
M. Hasenbusch,
\newblock Phys. Lett. B519 (2001) 177, hep-lat/0107019.

\bibitem{Wilson:1974sk}
K.G. Wilson,
\newblock Phys.Rev. D10 (1974) 2445.

\bibitem{Sheikholeslami:1985ij}
B. Sheikholeslami and R. Wohlert,
\newblock Nucl.Phys. B259 (1985) 572.

\bibitem{Luscher:1996sc}
M. L{\"u}scher, S. Sint, R. Sommer and P. Weisz,
\newblock Nucl.Phys. B478 (1996) 365, hep-lat/9605038.

\bibitem{soft:DDHMC}
M. {L\"uscher},
\newblock http://luscher.web.cern.ch/luscher/DD-HMC/index.html.

\bibitem{Hasenfratz:2001hp}
A. Hasenfratz and F. Knechtli,
\newblock Phys.Rev. D64 (2001) 034504, hep-lat/0103029.

\bibitem{soft:sparseLU}
T.A. Davis,
\newblock http://www.cise.ufl.edu/research/sparse/umfpack/.

\bibitem{Hasenfratz:2002ym}
A. Hasenfratz and A. Alexandru,
\newblock Phys.Rev. D65 (2002) 114506, hep-lat/0203026.

\bibitem{DelDebbio:2006cn}
L. Del~Debbio, L. Giusti, M. Luscher, R. Petronzio and N. Tantalo,
\newblock JHEP 0702 (2007) 056, hep-lat/0610059,
\newblock TeX source, 17 pages, figures included.

\bibitem{DelDebbio:2007pz}
L. Del~Debbio, L. Giusti, M. Luscher, R. Petronzio and N. Tantalo,
\newblock JHEP 0702 (2007) 082, hep-lat/0701009.

\bibitem{Cabibbo:1982zn}
N. Cabibbo and E. Marinari,
\newblock Phys.Lett. B119 (1982) 387.

\bibitem{Adler:1981sn}
S.L. Adler,
\newblock Phys.Rev. D23 (1981) 2901.

\bibitem{Petronzio:1990vx}
R. Petronzio and E. Vicari,
\newblock Phys.Lett. B245 (1990) 581.

\bibitem{Luscher:2007es}
M. L{\"u}scher,
\newblock JHEP 0712 (2007) 011, 0710.5417.

\bibitem{Wolff:2003sm}
ALPHA collaboration, U. Wolff,
\newblock Comput.Phys.Commun. 156 (2004) 143, hep-lat/0306017.

\bibitem{Finkenrath:2012cz}
J. Finkenrath, F. Knechtli and B. Leder,
\newblock (2012), 1211.1214.

\bibitem{Kennedy:2006ax}
A. Kennedy,
\newblock (2006), hep-lat/0607038.

\bibitem{Batrouni:1985jn}
G.G. Batrouni et~al.,
\newblock Phys. Rev. D32 (1985) 2736.

\bibitem{Davies:1987vs}
C.T.H. Davies et~al.,
\newblock Phys. Rev. D37 (1988) 1581.

\bibitem{Hasenfratz:2008fg}
A. Hasenfratz, R. Hoffmann and S. Schaefer,
\newblock Phys.Rev. D78 (2008) 014515, 0805.2369.

\bibitem{deForcrand:1998sv}
P. de~Forcrand,
\newblock Nucl.Phys.Proc.Suppl. 73 (1999) 822, hep-lat/9809145.

\bibitem{Duncan:1999xh}
A. Duncan, E. Eichten, R. Roskies and H. Thacker,
\newblock Phys.Rev. D60 (1999) 054505, hep-lat/9902015.

\end{thebibliography}
\bibliographystyle{h-elsevier}   %if you use h-elsevier.bst

\end{document}